\documentclass[fleqn,10pt]{wlscirep}
\usepackage[utf8]{inputenc}
\usepackage[T1]{fontenc}
\usepackage{caption}
\usepackage{subcaption}
\usepackage[linesnumbered,ruled,vlined]{algorithm2e}
\usepackage{xcolor} % For text coloring
\usepackage{amsmath}
\providecommand{\DIFadd}[1]{{\textcolor{black}{#1}}}  % New text in blue
\providecommand{\DIFdel}[1]{}  % Remove deleted text
\providecommand{\DIFaddbegin}{} %DIF PREAMBLE
\providecommand{\DIFaddend}{} %DIF PREAMBLE
\providecommand{\DIFdelbegin}{} %DIF PREAMBLE
\providecommand{\DIFdelend}{} %DIF PREAMBLE
\providecommand{\DIFaddFL}[1]{\DIFadd{#1}} %DIF PREAMBLE
\providecommand{\DIFdelFL}[1]{\DIFdel{#1}} %DIF PREAMBLE
\providecommand{\DIFaddbeginFL}{} %DIF PREAMBLE
\providecommand{\DIFaddendFL}{} %DIF PREAMBLE
\providecommand{\DIFdelbeginFL}{} %DIF PREAMBLE
\providecommand{\DIFdelendFL}{} %DIF PREAMBLE
\newcommand{\rev}[1]{#1}

\title{On the accurate computation of expected modularity in probabilistic networks}

\author[1,*]{Xin Shen}
\author[1]{Matteo Magnani}
\author[1]{Christian Rohner}
\author[2]{Fiona Skerman}
\affil[1]{InfoLab, Department of Information Technology, Uppsala University, Uppsala 75105, Sweden}
\affil[2]{Department of Mathematics, Uppsala University, Uppsala 75105, Sweden}

\affil[*]{xin.shen@it.uu.se}

\keywords{Modularity calculation, Probabilistic networks, Algorithms}

\begin{abstract}
\DIFaddbegin \DIFadd{
Modularity is one of the most widely used measures for evaluating communities in networks. In probabilistic networks, where the existence of edges is uncertain and uncertainty is represented by probabilities, the expected value of modularity can be used instead. However, efficiently computing expected modularity is challenging.
To address this challenge, we propose a novel and efficient technique ($\mathrm{FPWP}$) for computing the probability distribution of modularity and its expected value. }\DIFaddend %Instead of calculating over exponential \emph{possible worlds}, the technique calculates polynomial partitions, where each partition contains multiple possible worlds.  %DIF > 
\DIFaddbegin \DIFadd{%To accelerate the computation of partition probabilities, we leverage the Fourier transform. 
In this paper, we implement and compare our method and various general approaches for expected modularity computation in probabilistic networks. These include: (1) translating probabilistic networks into deterministic ones by removing low-probability edges or treating probabilities as weights, (2) using Monte Carlo sampling to approximate expected modularity, and (3) brute-force computation. We evaluate the accuracy and time efficiency of $\mathrm{FPWP}$ through comprehensive experiments on both real-world and synthetic networks with diverse characteristics. Our results demonstrate that removing low-probability edges or treating probabilities as weights produces inaccurate results, while the convergence of the sampling method varies with the parameters of the network. Brute-force computation, though accurate, is prohibitively slow. In contrast, our method is much faster than brute-force computation, but guarantees an accurate result.
}\DIFaddend %DIF > 
\end{abstract}
\begin{document}

\flushbottom
\maketitle
\thispagestyle{empty}

\section*{Introduction}
\DIFdelbegin %DIFDELCMD < 

%DIFDELCMD < %%%
\DIFaddbegin \DIFadd{Uncertainty is an inherent property when modelling a system as a network because of randomness of the system, inaccuracy of measurements, or their interpretation.
}\DIFaddend 
%It generates because of randomness of the system under study %DIF or inaccuracy of measurements. %DIF < system under study -- reality
System randomness can be found, for example, in computer networks where network links can be unreliable. Inaccuracy of measurements is ubiquitous, for example when estimating interaction probabilities in protein networks \cite{krogan2006global,danesh2022dgcu,yu2022stable} or assessing the existence of social relations in social networks \cite{farine2015estimating,liu2012reliable}. 
%To address these challenges, uncertainty %DIF < networks can be modeled %DIF < using %DIF < as \emph{probabilistic networks}, where associating each edge %DIF < are associated with a probability, \DIFdelbegin \DIFdel{representing }\DIFdelend \DIFaddbegin \DIFadd{which represent  
% MM representing the likelihood of its existence 
\DIFaddbegin \DIFadd{
Uncertainty is modelled associating each edge with a probability of existence, forming a \emph{probabilistic network}~\cite{frank1969shortest,kaveh2021defining,ceccarello2017clustering,danesh2023survey,hussain2021clustering,hussain2022clustering,liang_efficient_2020,danesh2021ensemble,pileggi2024cross}.
When the existence of all edges is certain, we talk of \emph{deterministic networks}\cite{banerjee2022survey,kollios2011clustering,danesh2023survey,halim2015clustering,ceccarello2017clustering,halim2019density}.}\DIFaddend

\DIFaddbegin \DIFadd{This paper focuses on the problem of calculating expected modularity in probabilistic networks.  Modularity is a  measure of the ratio of edges  falling  within  partitions  minus  approximately the expected ratio in an equivalent  network  with  edges  placed  at  random \cite{newman2006modularity}. Higher values of modularity often indicate that the input partitioning provides a good representation of the modules constituting the network. Therefore, despite some limitations of modularity as an objective function \cite{guimera2004modularity,mcdiarmid2020modularity,hanteer_unspoken_2020,peixoto2023descriptive}, modularity optimization \cite{chen2014community} has emerged as one of the most popular approaches for the analysis of deterministic networks.  }\DIFaddend

 \DIFaddbegin \DIFadd{
In probabilistic networks, modularity is not represented by a single value but by a distribution of modularity values arising from all combinations of edges in the network. These combinations are called \emph{possible worlds}, each having a different modularity and probability in general.  
There are no studies proposing specific algorithms to compute the expected modularity, which means that we currently have to rely on general methods. 
%To fill this gap, we implement approaches using several ways.
One way of computing expected  modularity in probabilistic networks is to calculate it over all possible worlds \cite{abiteboul1991representation,banerjee2022survey,danesh2023survey,ceccarello2017clustering},  and calculate its expected value. This approach is accurate; in fact, it is the only known approach producing the correct value of expected modularity, modulo numerical approximations. }\DIFaddend However, this approach is also computationally impractical as it involves calculating modularity $2^m$ times, where $m$ is the number of edges in the network.
%In \cite{yu2022stable}, the authors first 
Another approach is to sample $\theta$ 
%networks from all 
possible worlds, then calculate modularity for each sample
%d network 
and take the average value\cite{yu2022stable}.
\rev{This approach can be very fast, but prioritizes execution time over accuracy: one can speed up the execution by choosing a lower $\theta$, but at the cost of obtaining a potentially inaccurate result without a guaranteed approximation error.}
 Other general ways of handling 
%DIF < extending modularity to 
%DIF >  modularity to 
probabilistic networks, which can also be applied \DIFaddbegin \DIFadd{to }\DIFaddend modularity computation, include regarding edge probabilities as edge weights or setting a threshold to convert probabilistic networks to deterministic networks by removing edges with probability lower than the threshold. However, we show that these approaches are not appropriate for this specific task. \rev{Regarding probabilities as weights generally leads to wrong results, as we show in our experiments.}
%is non-meaningful as the  results  are different from those obtained using  edge probabilities. 
Setting a suitable threshold is also a complicated problem as it requires prior knowledge of the network structure and the chosen threshold greatly affects the resulting modularity \cite{kollios2011clustering}.
In fact, in this paper
%Section~\ref{sec:eval} 
we experimentally evaluate all the aforementioned approaches, 
and show that there is no known simple and general way of computing the expected value of modularity other than enumerating and computing modularity in all possible worlds, without \rev{a risk of} obtaining inaccurate results. 

Because of the limitations of existing approaches, in this paper we also introduce a new method for expected modularity computation. 
\DIFaddbegin \DIFadd{The novelty of this work is a new approach to exactly compute the modularity distribution without enumerating all possible worlds. }\DIFaddend Our method consists in partitioning the possible worlds so that for each partition we can (1) easily compute the expected value of modularity for the possible worlds in that partition, and (2) quickly compute the probability of the partition. The intuition behind our method is that, given a network and a clustering of its nodes, the value of modularity only depends on the number of edges inside and across communities, and not on the specific nodes incident to each edge. Therefore, we can group all possible worlds with the same numbers of in- and across-community edges into the same partition, obtaining that all possible worlds in each partition will have the same value of modularity. This would then also be the same as the expected modularity for the partition. This approach allows us to compute the correct value of expected modularity without having to compute modularity in all $2^m$ possible worlds, and to apply methods for 
the fast computation of probabilities from the Poisson Binomial distribution to the problem of expected modularity computation.
%, given that this distribution defines the probability of our partitions.
%
%\textbf{Our contribution.} We use an expression for modularity calculation based on communities instead of nodes. We exploit this expression to design efficient algorithms for calculating expected modularity in probabilistic networks. 
% MM
%\DIFdelbegin \DIFdel{Our algorithm }\DIFdelend \DIFaddbegin \DIFadd{The novelty of this paper is that we fill the gap of calculating expected modularity over probabilistic networks, except implementing basic computing methods, we also create a fast and accurate method $FPWP$. }\DIFaddend 
Our method reduces time complexity from exponential to polynomial compared to the brute-force approach.
Different from e.g. sampling, whose accuracy depends on the number of samples, our method always returns an accurate result, and its execution time does not depend on the edge probability distribution. Thanks to its ability to return an accurate result in polynomial time, our method also allows us to evaluate the traditional approaches used to analyse probabilistic networks but not providing guarantees on their accuracy, that is, sampling, thresholding, and weighting, when used to estimate expected modularity. 

%We also show that our method is robust with regard to different network structures, and study the factors influencing its running time.

%\textbf{Discussion.} Our algorithm outputs the expected modularity score based on an input set of communities. That is, it does not find communities in probabilistic networks. Using the proposed algorithm to find communities in probabilistic networks is material for future work.

%\textbf{Roadmap.} 
The rest of the paper is organized as follows: after reviewing the background
% in Section 2, we %propose our 
we 
present all the evaluated approaches, %algorithms in Section 3. 
including the new algorithm introduced in this work.
%In Section 4, 
Then we present a thorough simulation-based experimental evaluation on random 
%networks 
and real-world networks with different properties. We conclude with a summary and discussion of our results. % in Section 5.
The main notation used in the paper is summarized in Table~\ref{tab:notations}.

Please note that while this work is motivated by the importance of community detection, here we focus on the foundational problem of computing expected modularity given an input clustering (that is, modularity \emph{computation}) and not on how to use modularity to identify good clusterings (that is, modularity \emph{optimization}). In this paper a clustering is always given as an input of the algorithm, and expected modularity is computed for the input clustering.
\DIFaddbegin \DIFadd{
We also note that probabilistic networks and random graph models are related but distinct concepts. A probabilistic network represents a specific real-world system, and given two nodes in that network, an edge between them either exists or not 
in the real world.
%with an associated probability. 
Probabilities represent our ignorance with respect to the state of the real world. 
%The key difference lies in their treatment of probabilities. In probabilistic networks, probabilities represent the uncertainty or confidence in the existence of edges, but they do not imply that the edges are sampled. The network itself is typically treated as a whole, with fixed probabilities assigned to each edge. In contrast, random graphs are generated through a stochastic process, where edges are realized based on a probabilistic mechanism.
}\DIFaddend

\begin{table}[ht]
 \caption{Summary of notations %\mm{(1) $p$ is not a set (2) possibilities $\rightarrow$ probabilities (3) I find "communities in G" misleading, because the assignment to communities is not part of G. Also: please double-check that this is up-to-date with the current math in the paper, as we've been updating it.}
 }
    \centering
    \begin{tabular}{cc}
    %\begin{tabular}{@{}ll}
    \toprule
       probabilistic network  & $\mathcal{G}$ \\
       deterministic network  & $G$ \\
       number of nodes in $\mathcal{G}$  & $n$ \\
       number of edges in $\mathcal{G}$ & $m$\\
       set of nodes in $\mathcal{G}$& $V$\\
       set of edges in $\mathcal{G}$&$E$\\
       edge between node $i$ and node $j$ in $\mathcal{G}$  & $e_{ij}$\\
       probability of edge $e_{ij}$ & $p_{ij}$ \\
       edge between node $i$ and node $j$ in $G$  & $l_{ij}$\\
       set of possible worlds in $\mathcal{G}$ & $W =\{w_1,w_2,\dots,w_{2^m}\}$ \\
       probability of possible world $w$&$Pr(w)$\\
        set of communities in $\mathcal{G}$&$\mathcal{C}=\{c_1,c_2,\dots,c_k\}$\\
       number of communities in $\mathcal{C}$ &$k$\\
       set of nodes in community $c$&$V_{c}$\\
       set of edge probabilities in $\mathcal{G}$ & $P=\{p_1,p_2,\dots,p_{2^m}\}$\\

      % number of partition in $\mathcal{G}$ &$s$\\

       set of partitions of possible words in $\mathcal{G}$ &$D=\{d_1,d_2,\dots,d_s\}$\\%|d_i\cap d_j=\emptyset \wedge \sum_i d_i=W \wedge 1\leq i\leq j \leq s\}$\\
       %number of possible worlds in partition $d_i$ & $\mathcal{N}_i$\\
       %edge of node $i$ and node $j$ \mm{we only use this $l$ in the partitioning example, do we need it? (e.g. why not using $e$, or $(i,j)$)}& $l_{ij}$\\

       %edges completely within a community $c$&$e_c$\\
       %edges incompletely within a community $c$&$e_{c,\Bar{c}}$\\
       %edges completely outside a community $c$&$e_{\Bar{c}}$\\

       %probability of partition of possible worlds in $\mathcal{G}$&$Pr(D)$\\
       %modularity of a deterministic network &$Q$\\
       %maximum number of edges in $e_c$& $T_{c}$\\
       %maximum number of edges in $e_{c,\Bar{c}}$& $T_{c,\Bar{c}}$\\
       %maximum number of edges in $e_{\Bar{c}}$& $T_{\Bar{c}}$\\
       \bottomrule
    %\end{tabular}

    \label{tab:notations}
    \end{tabular}
\end{table}

%\IEEEPARstart{T}{his} demo file is intended to serve as a ``starter file''
%for IEEE Computer Society journal papers produced under \LaTeX\ using
%IEEEtran.cls version 1.8b and later.
% You must have at least 2 lines in the paragraph with the drop letter
% (should never be an issue)
%I wish you the best of success.

%\hfill mds

%\hfill August 26, 2015
\section*{Background}\label{sec: back}

%\subsection*{Probabilistic networks and relevant measures}
\subsubsection*{Probabilistic networks}

Consider a probabilistic network $\mathcal{G}=(V,E,p)$, where $V$ corresponds to the set of nodes in $\mathcal{G}$, $E$ represents the set of edges, and $p: E \rightarrow (0,1]$ is a function that assigns probabilities to edges. We use $e_{ij}$ to indicate the edge between nodes $i$ and~$j$, $p_{ij}$ as a shorthand for $p({e_{ij}})$, and we 
%assume that 
notate $|E|=m$. 

\DIFdelbegin \DIFdel{One can consider a probabilistic }\DIFdelend \DIFaddbegin \DIFadd{\emph{Possible worlds semantics} 
%has been recognized as a sound principle to define clustering on probabilistic networks\cite{kollios2011clustering,potamias2010k,abiteboul1991representation}. Generally, such a principle 
interprets a probabilistic network as a }\DIFaddend\DIFdelbegin \DIFdel{generative model for deterministic networks. }\DIFdelend \DIFaddbegin \DIFadd{set of deterministic networks, called possible worlds, each of which associated with its probability of being observed\cite{kollios2011clustering,potamias2010k,abiteboul1991representation}. That is, we
%DIF > One can consider a probabilistic network as a generative model for deterministic networks. 
}\DIFaddend 
%A deterministic network $G$ is generated by $\mathcal{G}$ connecting two nodes $i, j$ via an edge with probability $p_{ij}$. %Deterministic networks are an instance of probabilistic networks where $p_{ij} \in \{0,1\}$. MATTEO: I FOUND THIS CONFUSING, BECAUSE WE MIX DETERMINISTIC GRAPHS AS NON-PROBABILISTIC GRAPHS AND DETERMINISTIC GRAPHS AS POSSIBLE WORLDS. So I suggest commenting it.
%We write $G \subseteq \mathcal{G}$ to denote that $G$ is generated by $\mathcal{G}$, and we consider the edge probabilities independent of each other \cite{boonma2015reliable, parchas2015uncertain}. 
%We
\DIFdelbegin \DIFdel{use \emph{possible worlds} semantics \cite{abiteboul1991representation}, %DIF < which corresponds 
representing }\DIFdelend represent $\mathcal{G}$ with
%to 
the set $\{G=(V,E_G)\}_{E_G \subseteq E}$ of  all possible deterministic networks 
%generated by 
in $\mathcal{G}$ with their associated probabilities. 
\DIFaddbegin \DIFadd{As the edge probabilities are considered to be independent of each other\cite{boonma2015reliable, parchas2015uncertain}, the }\DIFaddend probability of observing such a set is:
\begin{align}
    \begin{split}
        Pr(G)=\prod_{e\in E_G}p(e)\prod_{e\in E \backslash E_G}(1-p(e)) \ .
        \label{poisson-binomial}
    \end{split}
\end{align}
There are $2^{m}$ distinct networks 
%that can be generated by 
\DIFaddbegin \DIFadd{in  }\DIFaddend
$\mathcal{G}$.

\subsubsection*{Modularity}
%\mm{Let's use a uniform terminology. In the following paragraph, I think \emph{part} and \emph{label} are synonyms with \emph{community}. If yes, let's just use community so it's clear we are referring to the same thing (or let's explicitly say that they are the same thing, if using \emph{part} and \emph{label} is important).}
A common way to identify communities in deterministic networks is to maximize the modularity score\rev{,} which rewards 
%partitions 
solutions capturing many of the edges within communities.
%and penalizes 
%partitions 
%solutions with few or large communities. % difference between edges within communities and edges between communities, which can be computed using modularity \cite{newman2004finding}. 
Giving a community labeling of vertices~$\textbf{x}$, the modularity for an undirected network is given by:
\begin{align}
    \begin{split}
        Q=\frac{1}{2M}\sum_i \sum_j(A_{ij}-\frac{k_ik_j}{2M})\delta(x_i,x_j) \ ,
        \label{e1}
    \end{split}
\end{align}
where $M$ is the number of edges, $A$ is the adjacency matrix, $k_i$ represents the degree of node $i$, $x_i$ is the label of node $i$, and $\delta(x_i,x_j)$ is the Kronecker delta function, which equals~1 when its arguments are the same and 0 otherwise. \DIFaddbegin \DIFadd{Modularity is a function of a network and a community assignment (that is, a partitioning of its nodes). However, it is common not to show its parameters and only write the function name ($Q$), to simplify the notation. }\DIFaddend

%\mmo{
%Modularity ($Q$) can also be expressed in the following form \cite{chen2014community}:
%\begin{align}
%    \begin{split}
%        Q=\sum_{c_i\in C}[\frac{|E_{c_i}^{in}|}{|E|}-(\frac{2|E_{c_i}^{in}|+|E_{c_i}^{out}|}{2|E|})^2],
%        \label{e1_2}
%    \end{split}
%\end{align}
%where $C$ is the set of all the communities, $c_i$ is a specific community in $C$, $|E_{c_i}^{in}|$ is the number of edges between nodes within community $c_i$, $|E_{c_i}^{out}|$ is the number of edges from the nodes in community $c_i$ to the nodes outside $c_i$, and $|E|=M$ is the total number of edges in the network.
%}

%\mme{I understand that \cite{chen2014community} uses the formula above. However, the formula that we use later is written differently, both in the notation (e.g. we replace $E_{c_i}^{in}$ with $e_c$) and in the denumerator (where we split $E$ into its three parts). This makes the whole thing heavy and inconsistent in notation. So my suggestion is to replace the above directly with the formula we use later, which is anyway just a different way of writing the one in \cite{chen2014community}.}

\begin{figure}[ht]
    \centering
    \includegraphics[width=.9\textwidth]{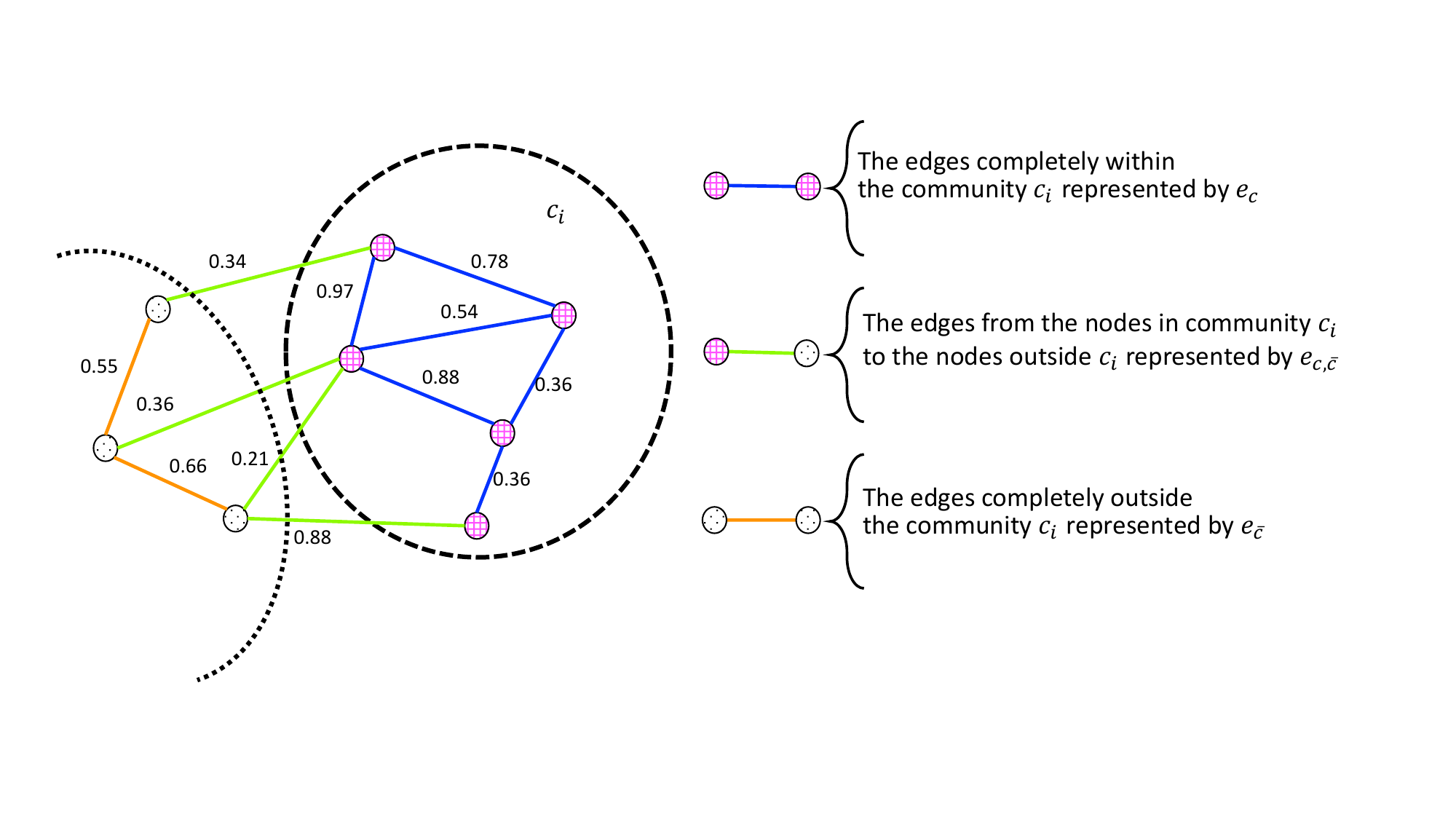}
    \caption{Edge partitions used in the community-based definition of modularity, for a community $c_i$.}
    \label{fig:edge_partitions}
\end{figure}

Modularity can also be expressed in the following form\cite{chen2014community}\DIFaddbegin \DIFadd{, which is the one used in our method}\DIFaddend:
\begin{align}
    \begin{split}
        Q=\sum_{c \in \mathcal{C}}\frac{|e_c|}{(|e_c|+|e_{c,\Bar{c}}|+|e_{\Bar{c}}|)}-(\frac{2|e_c|+|e_{c,\Bar{c}}|}{2(|e_c|+|e_{c,\Bar{c}}|+|e_{\Bar{c}}|)})^2 \ ,
        \label{eq:modularity_alt}
    \end{split}
\end{align} 
where $\mathcal{C}$ is the set of communities, $V_{c}$ is the set of nodes in community $c \in \mathcal{C}$, and 
\begin{align}
    \begin{split}
        e_c&=\{e_{ij}\ |\ i,j \in V_{c}\},\\
        e_{c,\Bar{c}}&=\{e_{ij}\ |\ (i \in V_{c} \land j \notin V_{c}) \lor (j \in V_{c} \land i \notin V_{c})\}, \\
        e_{\Bar{c}}&=\{e_{ij}\ |\ i,j \notin V_{c}\}.
        \label{def_of_e}
    \end{split}
\end{align}
An example of $e_c$, $e_{c,\Bar{c}}$, and $e_{\Bar{c}}$ for a community $c_i$ is presented in Figure~\ref{fig:edge_partitions}.

%_______________________________
\subsubsection*{Expected modularity}
A deterministic network $G_w$ 
%generated by 
\DIFaddbegin \DIFadd{in }\DIFaddend
$\mathcal{G}$ can be  considered as one possible world. From Eq.~\ref{poisson-binomial}, we know the probability of each possible world in $\mathcal{G}$% \rev{, that is $Pr(w)$}
. Let $\textbf{Q}$ be the discrete distribution of $Q$ on the $2^m$ possible worlds. The  expected value of~$\textbf{Q}$ is:

\begin{align}
    \begin{split}
        E(\textbf{Q})=\sum_{w=1}^{2^m} Q_w Pr(w),
        \label{e4}
        \end{split}
\end{align}
where  $Q_w$ is the modularity value of deterministic network~$G_w$ (that is, possible world $w$). %DIF > Therefore, $Pr(w)$ can be computed using Eq.~\ref{poisson-binomial}.

\subsubsection*{Entropy ratio}
%\mm{We should say why we need entropy, and also (I suggest) be even more explicit about why we normalise --- are we doing it to be able to compare networks with different sizes, or to understand what the values mean, or? and also that the range is [0,1].}
The entropy \cite{parchas2018uncertain, kaveh2022modelling} $H(\mathcal{G})$ of a probabilistic network $\mathcal{G}$ is defined as the joint entropy of its edges $H(e)$ for all $e \in E$. 
Here we use 
%different 
entrop\DIFaddbegin \DIFadd{y }\DIFaddend to represent the different levels of uncertainty of a network. The larger entropy, the more uncertainty. Due to the edge independence, the formula of entropy is $H(\mathcal{G})=-\sum_{i<j}p_{ij}\log p_{ij}-\sum_{i<j}q_{ij}\log q_{ij}$, where $q_{ij}=1-p_{ij}$. To easily compare networks with different sizes, we normalize such entropy by dividing by the number of edges. We call this \emph{entropy ratio}, where entropy ratio =$\frac{|\mathcal{H}(\mathcal{G})|}{m}$, and its range is [0,1].

\section*{Computation methods}

\subsubsection*{Brute-force}
The brute-force method to \DIFaddbegin \DIFadd{compute }\DIFaddend expected modularity \DIFaddbegin \DIFadd{directly uses Eq.~\ref{e4}, }\DIFaddend
%of probabilistic network $\mathcal{G}$ is by dividing such probabilistic network into $2^m$ possible worlds \cite{ghosh2007routing}, 
calculating the modularity on every possible world and their expected value\DIFaddbegin\DIFadd{\cite{ghosh2007routing}}\DIFaddend. Therefore, the time complexity of this method is at least exponential on the size of the network. %\mm{Should this be $\Theta(2^m)$, as suggested by Fiona?} \Xin{yes, here brute force means every time it takes $2^m$, it is average. }
While impractical, this method returns the correct value of expected modularity, and can thus be used on small networks to evaluate the accuracy of other approaches.

\subsubsection*{Sampling}
%\mm{I think "classic" is either ambiguous (if there are different types of sampling, we should use the specific name of the type we use) or unnecessary (if there is only one "reasonable" definition of sampling \emph{from a probability distribution}, which I think is the case? In which case we should just say: Sampling method, or Sampling-based method).}
%We use a
The sampling method called 
Monte Carlo estimator%, which is introduced in 
\cite{wing1964analysis}
%, which is also been used in \mbox{%DIFAUXCMD
%\cite{potamias2010k,li2015recursive}}\hskip0pt%DIFAUXCMD
%Specifically,
%the MC estimator  
first generates %$\theta$ 
$\theta$ samples from the probability distribution over the possible worlds, then calculates modularity for each sample
%d network 
\DIFaddbegin \DIFadd{$t$ (which we notate $Q_t$) }
and computes the average modularity value:
\begin{align}
    \begin{split}
        Q=\frac{1}{\theta}\sum_{t=1}^\theta Q_t \ .
        \label{aver_mod}
    \end{split}
\end{align}
Once the parameter $\theta$ has been specified, creating samples from the ensemble is straightforward. The detailed algorithm is depicted in Algorithm~\ref{sampling}. Note that we do not use Naive Monte Carlo sampling, that is known to have a higher variance \cite{li_recursive_2016}.
%\mm{About the algorithm: (1) Line 1 sounds like we already have $\theta$ sampled networks. We should maybe just iterate from $z \leftarrow$ 1 to $\theta$, create a network inside the loop, add the edges to this network, and add the network to the result (maybe a variable created before the loops) after the internal loop? (2) Line 2, shouldn't we iterate over E? (3) Line 4: is "undirected" necessary --- aren't we only working with undirected edges in this paper? I ask because it's more readable if each instruction is on a single line. Same for $r_{ij}$: is $ij$ needed as we regenerate r every time?}
%\Xin{I think it is undirected network as our modularity formula is undirected. And we generate r every time, each r is corresponded to one specific node pair, so I think it might necessary.}
%\mmo{
%\begin{enumerate}
%    \item for each pair of nodes $i$, $j$, draw a uniformly random number $r_{ij}$
%    \item if $r_{ij} \leq P_{ij}$, then add the undirected edge ($i$,$j$) to sample network $z$
%    \item repeat step 1) and step 2) until $z=\theta$   
%\end{enumerate}
%}

\IncMargin{1em}
\begin{algorithm}
\SetKwData{Left}{left}\SetKwData{This}{this}\SetKwData{Up}{up}
\SetKwFunction{Union}{Union}\SetKwFunction{FindCompress}{FindCompress}
\SetKwInOut{Input}{input}\SetKwInOut{Output}{output}
\Input{The number of networks to sample $\theta$; nodes $V$; edges $E$; edges probabilities $p$}
\Output{Set of networks $Z$}
\BlankLine
Initialize $Z$\;
\For{$z \leftarrow$ 1 \KwTo $\theta$}{
Create an empty network $g$ only with nodes $V$\;
\For{each pair of nodes $i,j \in E$}{
Draw a uniformly random number $r_{ij} \in [0,1]$\;
if $r_{ij} \leq p_{ij}$, then add the edge ($i$,$j$) to network $g$\;
}
add $g$ to $Z$\;
}
\Return{$Z$}\;
\caption{Sampling method}\label{sampling}
\end{algorithm}\DecMargin{1em}

%\textbf{Random sampling.} In  random sampling method, we randomly choose $T$ samples from all possible worlds with replacement. Using Eq.~\ref{aver_mod} and let $\theta=T$. Similarly, this method time complexity is also exponential according to the number of samples, which is $O(2^T)$.

%The detailed algorithm is depicted in Algorithm~\ref{pro_quota}.

%\IncMargin{1em}
%\begin{algorithm}
%\SetKwData{Left}{left}\SetKwData{This}{this}\SetKwData{Up}{up}
%\SetKwFunction{Union}{Union}\SetKwFunction{FindCompress}{FindCompress}
%\SetKwInOut{Input}{input}\SetKwInOut{Output}{output}
%\Input{Probability of edges $p$; Number of samples $z$; Number of edges $E$}
%\Output{Samples of possible worlds $S_{pw}$}
%\BlankLine
%Initialize $S_{pw}$\;
%$S_{pw} \leftarrow []$\;
%\While{the number of samples in $S_{pw}$ $\leq z$}{
%$pw \leftarrow []$\;
%\For{$j\leftarrow 0$ \KwTo $(E-1)$}{\label{forins}
%$r_{ij} \leftarrow random(0,1)$\;
%\If(\tcp*[f]{$p_{ij}$ is the probability of node pair $i$ and $j$}){$r_{ij} < p_{ij}$}{
%$pw.insert(1)$}
%\lElse{$pw.insert(0)$}

%}
%$S_{pw}.insert(pw)$\;

%}
%\Return{$S_{pw}$}\;
%\caption{Classic Sampling}\label{pro_quota}
%\end{algorithm}\DecMargin{1em}

\subsubsection*{Thresholding}   
%\mm{Why "methods" (plural)?}
Thresholding is a simple \DIFaddbegin\DIFadd{and general }\DIFaddend method 
%currently 
used in community detection in probabilistic networks. Its purpose %of using threshold methods 
is to transform probabilistic networks into deterministic networks, so that existing methods for deterministic networks can be applied. After setting a threshold, we remove edges whose probability is lower than the threshold and consider the others as deterministic.
%\subsubsection*{Entropy ratio}
%The entropy \cite{parchas2018uncertain, kaveh2022modelling} $H(\mathcal{G})$ of an uncertain network $\mathcal{G}$ is defined as the joint entropy of its edges $H(e)$ for all $e \in E$. Due to the edge independence, $H(\mathcal{G})=-\sum_{i<j}p_{ij}\log p_{ij}-\sum_{i<j}q_{ij}\log q_{ij}$, where $q_{ij}=1-p_{ij}$. We normalize such entropy by dividing by the number of edges. We called it entropy ratio, where entropy ratio =$\frac{|\mathcal{H}(G)|}{m}$.
%We introduce the absolute value of entropy in network construction, the absolute value of entropy definition is shown in Equation \ref{e7}.
%\begin{align}
%    \begin{split}
%         \mathcal{H}(G)&=-\sum_{i<j}p_{ij}\log p_{ij}-\sum_{i<j}(1-p_{ij})\log(1-p_{ij}).\\

 %   \label{e7}
 %   \end{split}
%\end{align}
%Let $|\mathcal{H}(G)|$ divided by the number of edges $m$ in probabilistic network, we normalize such entropy from 0 to 1, where entropy ratio =$\frac{|\mathcal{H}(G)|}{m}$.

%\section*{Proposed algorithms}\label{sec: alg}

%In this section we introduce two algorithms to compute expected modularity, exactly and approximately. 

\subsection*{Possible-World Partitioning (PWP)}

The Possible-World Partitioning for Expected Modularity ($\mathrm{PWP}$) algorithm, that we introduce in this work, 
% 
%algorithm is a variation of the brute-force method where instead of iterating through all possible worlds we 
groups the possible worlds into partitions so that all possible worlds inside the same partition have the same value of modularity. \DIFaddbegin \DIFadd{For a given community assignment $\mathcal{C}$ and }\DIFaddend 
using the alternative definition of modularity in Eq.~\ref{eq:modularity_alt}, expected modularity can be rewritten as:
\begin{align}
    \begin{split}
        E(\textbf{Q})=\sum_{w=1}^{2^m} \sum_{c \in \mathcal{C}} Q_c^w Pr(w) \ ,
        \label{e4_3}
        \end{split}
\end{align}
where $w$ indicates a possible world corresponding to deterministic network $G_w$, $Pr(w)$ is the probability of this possible world, % $w$,
\begin{align*}
Q_c^w = \frac{|e_c^w|}{(|e_c^w|+|e_{c,\Bar{c}}^w|+|e_{\Bar{c}}^w|)}-(\frac{2|e_c^w|+|e_{c,\Bar{c}}^w|}{2(|e_c^w|+|e_{c,\Bar{c}}^w|+|e_{\Bar{c}}^w|)})^2 \ ,
\end{align*}
and $e_c^w$, $e_{c,\Bar{c}}^w$, and $e_{\Bar{c}}^w$ refer to $e_c$, $e_{c,\Bar{c}}$, and $e_{\Bar{c}}$ in $G_w$. %possible world $w$.

After rearranging the two sums:
\begin{align}
    \begin{split}
        E(\textbf{Q})&=\sum_{c \in \mathcal{C}} \sum_{w=1}^{2^m}  Q_c^w Pr(w) \ ,
        \label{e4_2}
        \end{split}
\end{align}
we can partition the $2^m$ possible worlds 
%so that in each partition the values of $|e_c^w|$, $|e_{c,\Bar{c}}^w|$, and $|e_{\Bar{c}}^w|$ are constant. 
\DIFaddbegin \DIFadd{
in a way that allows us to process all the possible worlds in each partition without iterating over them. In particular, the
key idea of our approach to reduce %the 
computational complexity 
%to calculate the expected modularity 
is to 
%partition all possible worlds $w$ 
define these partitions
so that the possible worlds within the same partition have the same modularity. %Specifically, we look for 
We do this by defining partitions whose possible worlds have the same number of edges in $e_c^w$, $e_{c,\Bar{c}}^w$, and $e_{\Bar{c}}^w$, respectively. Note that these sets are disjoint and in general have different sizes. 
We notate $d^{xyz}$ the partition containing all possible worlds $w$ where $|e_c^w| = x$, $|e_{c,\Bar{c}}^w| = y$, and $|e_{\Bar{c}}^w| = z$. This partition contains $\binom{T_x}{x} \cdot \binom{T_y}{y} \cdot \binom{T_z}{z}$ possible worlds, where $T_x = |e_c|$,
$T_y = |e_{c,\Bar{c}}|$, and
$T_z = |e_{\Bar{c}}|$ are the (maximum) number of edges in the three parts, respectively. We can then write:
}\DIFaddend
\begin{align}
    \begin{split}
        \DIFaddbegin \DIFadd{E(\textbf{Q})=\sum_{c \in \mathcal{C}} \sum_{x=0}^{T_x} \sum_{y=0}^{T_y} \sum_{z=0}^{T_z} Q_c^{xyz} Pr( d^{xyz}}\DIFaddend ) \ ,
        \label{exp_mod_partitioned}
        \end{split}
\end{align}
%\Xin{I changed x=1,y=1,z=1 to =0.}
%\DIFaddbegin \DIFadd{where $d^{xyz}$  indicates the partition (that is, the set) of possible worlds where $|e_c^w| = x$, $|e_{c,\Bar{c}}^w| = y$, and $|e_{\Bar{c}}^w| = z$. }\DIFaddend 
\DIFaddbegin\DIFadd{where  $Q_c^{xyz}$ is the constant value of $Q_c^w$ in all possible worlds $w \in d^{xyz}$, that is:}\DIFaddend
\begin{align}
\begin{split}
Q_c^{xyz} = \frac{x}{(x+y+z)}-(\frac{2x+y}{2(x+y+z)})^2 \ .
\label{Qc_xyz}
\end{split}
    \end{align}
%
%When $x=y=z=0$, the corresponding network has no edges and $Q_c^{xyz}=0$.
Notice that computing expected modularity using Eq.~\ref{exp_mod_partitioned} we only need to iterate over $|\mathcal{C}|T_xT_yT_z$ terms instead of $2^m$, and for each term the value of $Q_c^{xyz}$ can be computed in constant time. 
The probability of partition $d^{xyz}$ can be expressed as a product of probabilities, because $e_c^w$, $e_{c,\Bar{c}}^w$, and $e_{\Bar{c}}^w$ are always disjoint sets:
\begin{align}
\begin{split}
\DIFaddbegin \DIFadd{Pr(d^{xyz}) 
=Pr(|e_c^w| = x \land |e_{c,\Bar{c}}^w| = y \land |e_{\Bar{c}}^w| = z)
= Pr(|e_c^w|=x) \cdot Pr(|e_{c,\Bar{c}}^w|=y) \cdot Pr(|e_{\Bar{c}}^w|=z) \ .}\DIFaddend
\label{abs}
\end{split}
\end{align}

\DIFaddbegin \DIFadd{These probabilities can be computed using the definition of Poisson Binomial distribution. 
%here $A$ is a set from $F_x$ or $F_y$ or $F_z$, where $F_{x/y/z}$ is the set of all subsets of $\delta_{x/y/z}$ of size  ${x/y/z}$, 
In the following equations, $A$ is an element of $F_x$, $F_y$ or $F_z$, 
where $F_{x}$ is the set of all subsets of $e_c$ of size  $x$, 
$F_{y}$ is the set of all subsets of $e_{c,\Bar{c}}$ of size  $y$,
$F_{z}$ is the set of all subsets of $e_{\Bar{c}}$ of size  $z$.
}\DIFaddend 
%and the probabilities on the edges in $e_c^w$, $e_{c,\Bar{c}}^w$, and $e_{\Bar{c}}^w$ are, %\mm{Add formula, to make the paper self-contained.}
%\begin{align}
%    \begin{split}
%        Pr(xyz)=\prod_{\varrho \in  \{e_c,e_{c,\Bar{c}},e_{\Bar{c}}\}} \prod_{e\in \varrho^w}p(e)\prod_{e\in \varrho \backslash \varrho^w}(1-p(e)),
%        \label{mul_poisson-binomial}
%    \end{split}
%\end{align}
\begin{align}
    \begin{split}
        Pr(|e_c^w|=x)=
        \sum_{A \in F_x }\prod_{e\in A}p(e)\prod_{e\in  e_c\backslash A}(1-p(e)) \ ,
        \label{pr_x}
    \end{split}
\end{align}

\begin{align}
    \begin{split}
        Pr(|e_{c,\Bar{c}}^w|=y)=
        \sum_{A \in F_y }\prod_{e\in A}p(e)\prod_{e\in  e_{c,\Bar{c}}\backslash A}(1-p(e)) \ ,
        \label{pr_y}
    \end{split}
\end{align}
\begin{align}
    \begin{split}
        Pr(|e^w_{\Bar{c}}|=z)=
        \sum_{A \in F_z }\prod_{e\in A}p(e)\prod_{e\in  e_{\Bar{c}}\backslash A}(1-p(e)) \ .
      \label{pr_z}
   \end{split}
\end{align}
%Then the joint probability $Pr(xyz)$ is,
% \begin{align}
%     \begin{split}
%         Pr(xyz)=\prod_{\varrho \in  \{x, y, z\}} 
%         \sum_{A \in F_\varrho }\prod_{e\in A}p(e)\prod_{e\in \delta_\varrho \backslash A}(1-p(e)) \ ,
%         \label{mul_poisson-binomial}
%     \end{split}
% \end{align}

%     \begin{equation*}
%     \delta_\varrho= \left\{ 
%     \begin{array}{rcl}
% e_c& \mbox{for}
% & \varrho=x\\ e_{c,\Bar{c}} & \mbox{for} &  \varrho=y\\
% e_{\Bar{c}}& \mbox{for} & \varrho=z
% \end{array}\right. \end{equation*}
%FIONA: changed $ ... $ to \[ ... \]
%
%where $\varrho^w$ refers to $e_c^w$, $e_{c,\Bar{c}}^w$, and $e_{\Bar{c}}^w$ according to $\varrho \in  \{e_c,e_{c,\Bar{c}},e_{\Bar{c}}\}$.
%where $F_\varrho$ is the set of all subsets of $\delta_\varrho$ of size  $\varrho$. %For example, $F_x$ will contain $\frac{T_x !}{(T_x -x)!x!}$ elements. 
%The probabilities for disjoint sets $e_c^w$, $e_{c,\Bar{c}}^w$ and $e_{\Bar{c}}^w$ are,

\begin{figure}[ht] %{.65\textwidth}
    \centering
        \includegraphics[clip,width=.65\textwidth]{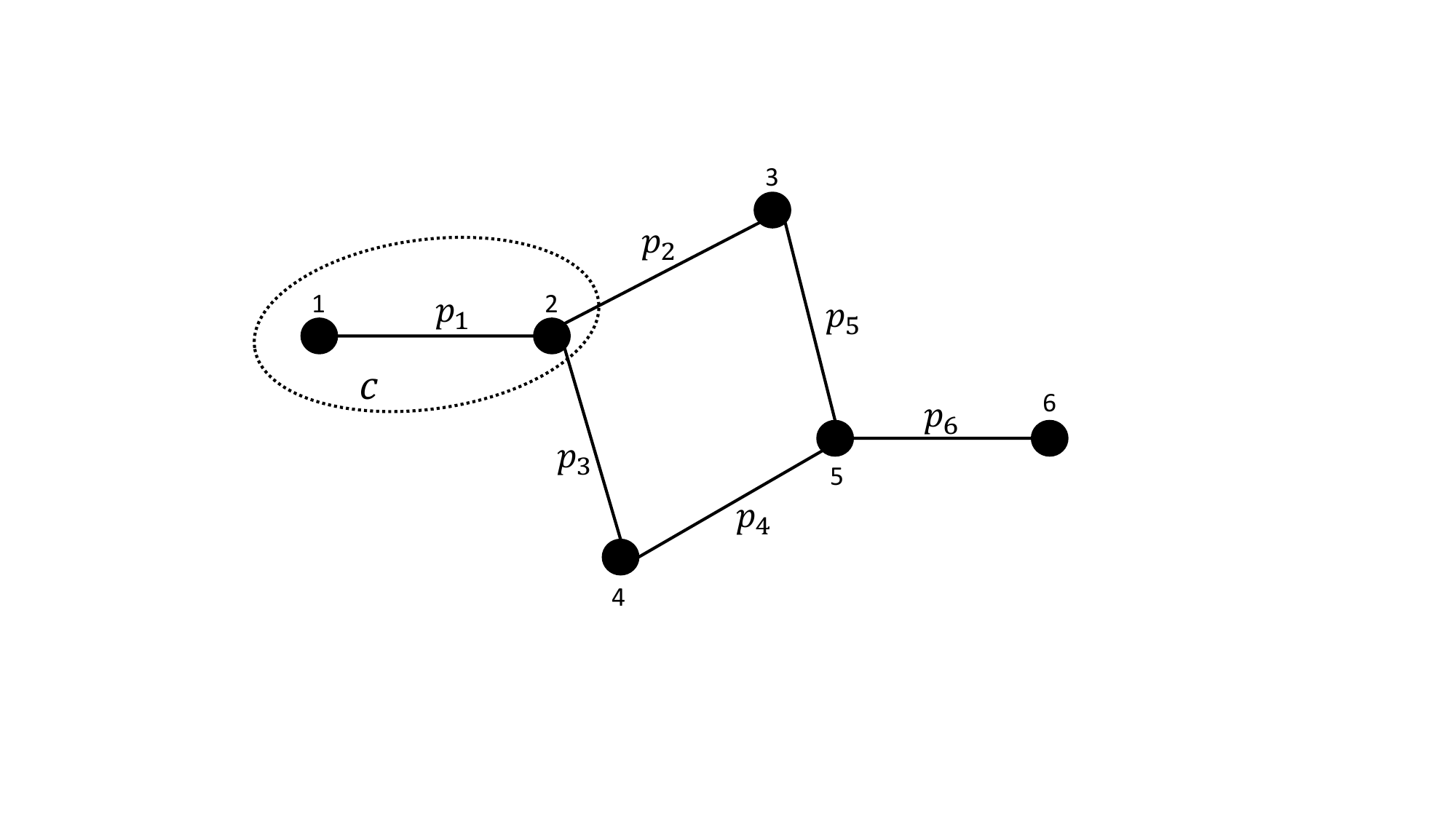}
        \caption{A probabilistic network with a community $c$.}
        \label{p3}
\end{figure}

\begin{figure}[ht] %{.54\textwidth}
    \centering
  \includegraphics[clip,width=.9\textwidth]{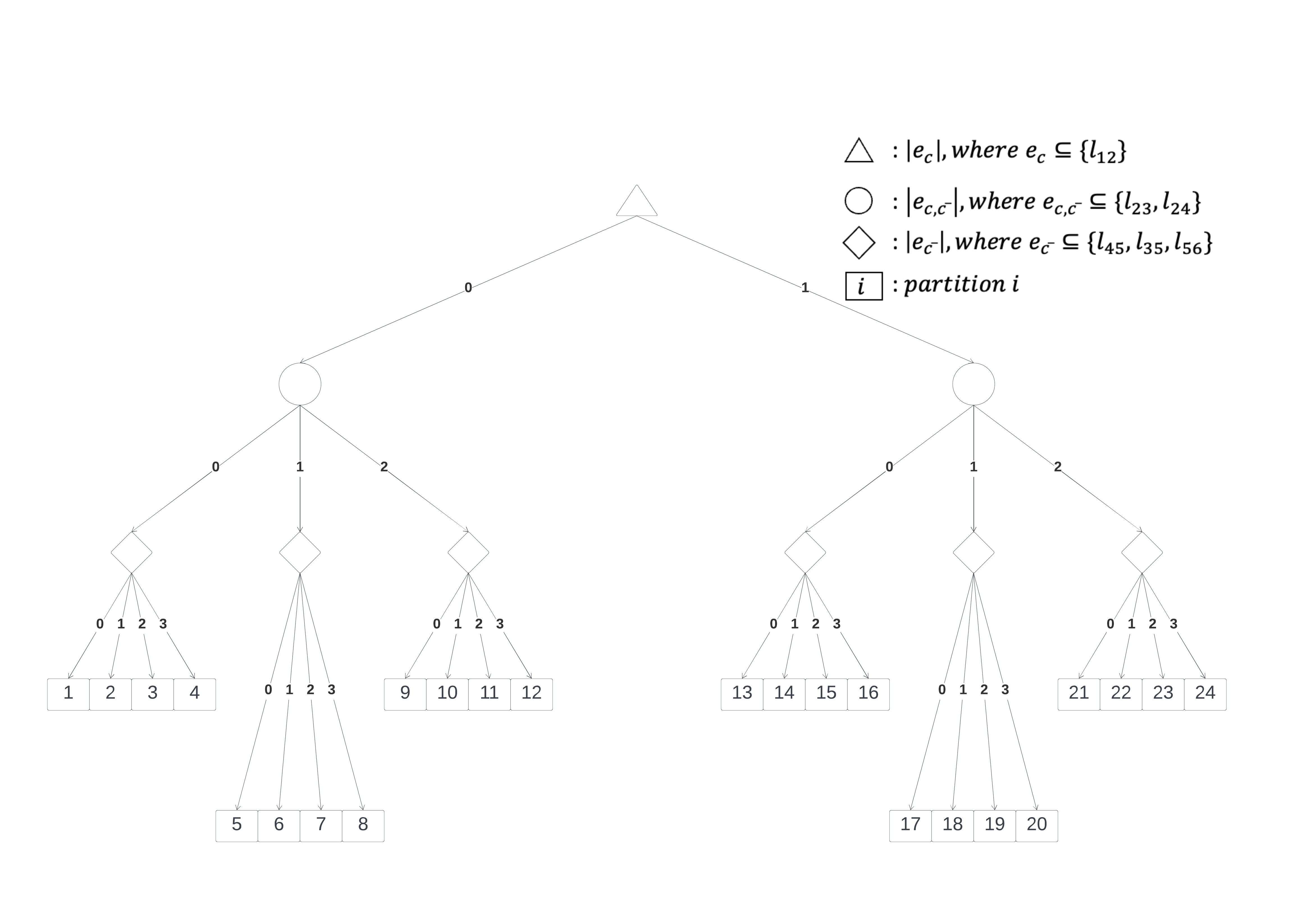}%
  \caption{A tree enumerating all partitions of possible worlds defined by community $c$. %\mm{This is really a small and ultra-picky comment, but the figure can be improved: the leaves of the tree (and maybe also the numbers) are not perfectly aligned, the lines sometimes reach the rectangle sometimes not, \dots}
  }
  \label{p1}
\end{figure}

\begin{figure}[ht] %{.7\textwidth}
    \centering
  \includegraphics[clip,width=.9\textwidth]{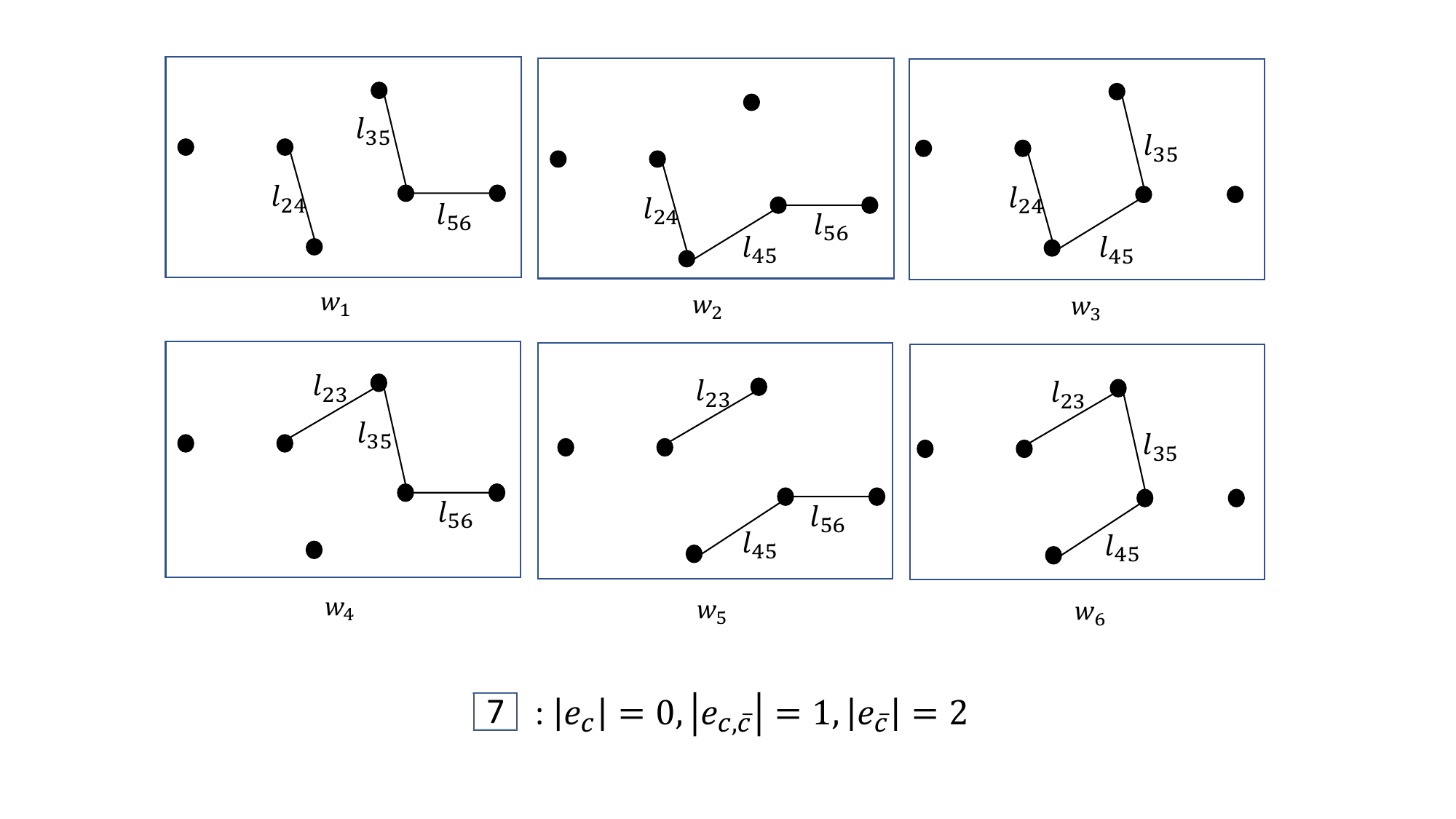}%
  \caption{All possible worlds in partition 7. $|e_c^w|, |e_{c,\Bar{c}}^w|$, and $|e_{\Bar{c}}^w|$ are constant inside the partition.}
  \label{p2}
\end{figure}

As an example of how the partitions are defined, consider the probabilistic network in Figure~\ref{p3}. Figure~\ref{p1} shows a tree whose leaves enumerate all partitions \DIFaddbegin \DIFadd{$d^{xyz}$ }\DIFaddend defined by community $c$. The two branches from the top triangle represent respectively all the possible worlds where $e_c^w$ contains no edges ($x=0$) and all the possible worlds where $e_c^w$ contains one edge ($x=1$), in this case edge $l_{12}$.
%and can be split into two cases:  exist in $e_c$ or one edge exists in $e_c$. 
As we move down the tree, the circles in the second level represent all the possible worlds where  $e_{c,\Bar{c}}^w$ contains no edges ($y=0$), one edge ($y=1$, in particular $l_{23}$ or $l_{24}$), and two edges ($y=2$, in this case both $l_{23}$ and $l_{24}$). The same with the diamonds in the third level representing possible worlds with a specific number of edges in $e_{\Bar{c}}^w$. The squares at the bottom of the figure represent all partitions. Figure~\ref{p2} shows all possible worlds in partition 7, where $|e_c^w|=0, |e_{c,\Bar{c}}^w|=1, |e_{\Bar{c}}^w|=2$.

$\mathrm{PWP}$ 
is detailed in Algorithm~\ref{apwp}. In lines 6, 8 and 10, we use Eq.~\ref{pr_x}, Eq.~\ref{pr_y} and Eq.~\ref{pr_z} to calculate $Pr(|e_c^w|=x)$,  $Pr(|e_{c,\Bar{c}}^w|=y)$ and $Pr(|e_{\Bar{c}}^w|=z)$ separately.

%\textbf{\rev{Computational complexity:}} 
%DIF < DIF > the worst-case time complexity of $\mathrm{APWP}$ is polynomial, and depends on two parts. One part is the loop time over all partitions, whose time complexity is $O(|E|^3)$. In the worst case, the number of edges in $e_c$, $e_{c,\Bar{c}}$ and $e_{\Bar{c}}$ are similar to each other. The second part is the approximation time for the  probability of each partition, whose time complexity is $O(|E|^2)$. In the worst case, the number of edges in one of $e_c$, $e_{c,\Bar{c}}$ and $e_{\Bar{c}}$ is much larger than in the other two. We only calculate $pr(x) Pr(y) and Pr(z)$ once for each partitions.
%DIF > DIF > the worst-case time complexity of $\mathrm{FPWP}$ is polynomial, and depends on two parts. One part is the loop time over all partitions, whose time complexity is $O(|E|^3)$. In the worst case, the number of edges in $e_c$, $e_{c,\Bar{c}}$ and $e_{\Bar{c}}$ are similar to each other. The second part is the approximation time for the  probability of each partition, whose time complexity is $O(|E|^2)$. In the worst case, the number of edges in one of $e_c$, $e_{c,\Bar{c}}$ and $e_{\Bar{c}}$ is much larger than in the other two. We only calculate $pr(x) Pr(y) and Pr(z)$ once for each partitions.

While $\mathrm{PWP}$ can already be considered a usable algorithm, because it reduces the number of modularity computations from exponential to polynomial, its worst-case time complexity 
%of $\mathrm{PWP}$
% 
is still $O(km2^m)$. We can break the analysis down into two parts.
Lines 3 to 10 concern the exact calculation of probabilities for each partition. The time complexities of $Pr(x)$, $Pr(y)$, and $Pr(z)$ are $O(2^m)$. From Equation~\ref{exp_mod_partitioned}, we can see that $x$ ranges from 1 to $T_x$, $y$ ranges from 1 to $T_y$, and $z$ ranges from 1 to $T_z$. $T_x$, $T_y$, and $T_z$ sum to $m$, so the overall complexity is $O(m2^m)$. Lines 11 to 15 involve looping over all partitions, which has a time complexity of $O(m^3)$. Notice that in the worst-case scenario, the number of edges in $e_c$, $e_{c,\Bar{c}}$, and $e_{\Bar{c}}$ are similar to each other. Here the number of edges in $e_c$, $e_{c,\Bar{c}}$, and $e_{\Bar{c}}$ have the same upper bound $m$. Both parts are inside the community loop on line 2, so the time complexity of the whole algorithm is $O(k(m2^{m} + m^3)) = O(km2^{m})$, where  $k$ is the number of communities. Notice that the time complexity solely depends on the number of communities and edges, and remains unaffected by the assignment of edge probabilities.

%\begin{figure}[ht]
%\captionsetup{justification=centering}

%\caption{An example demonstrating how, for a given community, all possible worlds of a probabilistic network can be partitioned so that $|e_c^w|, |e_{c,\Bar{c}}^w|$, and $|e_{\Bar{c}}^w|$ are constant inside each partition.}
%    \label{fig:par_pw}
%\end{figure}

\DIFaddbegin \DIFadd{
\subsection*{Fast/Fourier Possible-World Partitioning (FPWP)}
While PWP practically only can be used in small networks, its partitioning approach, combined with the Poisson Binomial distribution, enables a further reduction in computational complexity. The Poisson Binomial distribution can be expressed in closed-form using the Discrete Fourier Transform~\cite{fernandez2010closed}; 
\begin{align}
   \begin{split}
       \DIFadd{Pr(X=k)=\frac{1}{n+1}\sum_{l=0}^{n} C^{-lk}\prod_{i=1}^{n} (1+(C^l-1)p_{i}^k) \
   }\end{split}
\end{align}
where $C=exp(j\frac{2\pi}{n+1})$ and $j=\sqrt{-1}$.
}

\DIFaddbegin \DIFadd{
This leads to a version of our method that we call Fast/Fourier Possible World Partitioning for Expected Modularity ($\mathrm{FPWP}$), replacing equations~\ref{pr_x}, \ref{pr_y}, \ref{pr_z} by the following equations~\ref{appro_prx}, \ref{appro_pry}, \ref{appro_prz}, respectively:}   
\begin{align}
    \begin{split}
        Pr(|e_c^w|=x)
        &\DIFdelbegin \DIFdel{\approx  }\DIFdelend \DIFaddbegin \DIFadd{=  }\DIFaddend \frac{1}{T_{x}+1}\sum_{l=0}^{T_{x}} C^{-lx}\prod_{\alpha=1}^{T_{x}} (1+(C^l-1)p_{\alpha}^x) \ ,
        \label{appro_prx}
    \end{split}
\end{align}
\begin{align}
    \begin{split}
        Pr(|e_{c,\Bar{c}}^w|=y)
        &\DIFdelbegin \DIFdel{\approx  }\DIFdelend \DIFaddbegin \DIFadd{=  }\DIFaddend \frac{1}{T_{y}+1}\sum_{l=0}^{T_{y}} C^{-ly}\prod_{\alpha=1}^{T_{y}} (1+(C^l-1)p_{\alpha}^y) \ ,
        \label{appro_pry}
    \end{split}
\end{align}
\begin{align}
    \begin{split}
        Pr(|e_{\Bar{c}}|=z)
        &\DIFdelbegin \DIFdel{\approx  }\DIFdelend \DIFaddbegin \DIFadd{= }\DIFaddend \frac{1}{T_{z}+1}\sum_{l=0}^{T_{z}} C^{-lz}\prod_{\alpha=1}^{T_{z}} (1+(C^l-1)p_{\alpha}^z) \ .
        \label{appro_prz}
    \end{split}
\end{align}

\DIFdelbegin \DIFdel{$\mathrm{APWP}$ 
}\DIFdelend \DIFaddbegin \DIFadd{$\mathrm{FPWP}$ 
}\DIFaddend % 
is \rev{also} detailed in Algorithm~\ref{apwp}. In \rev{this case, differently from $\mathrm{PWP}$,} in lines 6, 8 and 10 we use Eq\rev{.}~\ref{appro_prx}, Eq\rev{.}~\ref{appro_pry} and Eq\rev{.}~\ref{appro_prz} to calculate $Pr(|e_c^w|=x)$,  $Pr(|e_{c,\Bar{c}}^w|=y)$ and $Pr(|e_{\Bar{c}}^w|=z)$ separately.
%___________________
%\PAR
\begin{algorithm}[tbp]
\SetKwData{Left}{left}\SetKwData{This}{this}\SetKwData{Up}{up}
\SetKwFunction{Union}{Union}\SetKwFunction{FindCompress}{FindCompress}
\SetKwInOut{Input}{input}\SetKwInOut{Output}{output}
\Input{A set of communities;  Edges: $E$; Edge probabilities: $p$ }
\Output{Expected modularity: $Sum$}
\BlankLine
$Sum=0$;\\
\For{$c\leftarrow 1$ \KwTo $k$}{
Initialize $P_c$, $P_{c,\Bar{c}}$ and $P_{\Bar{c}}$ as empty arrays;\\
%\tcp*[r]{$P_c$ collects $Pr(|e_c^w|=x)$, and $P_{c,\Bar{c}}$, $P_{\Bar{c}}$ collect $Pr(|e_{c,\Bar{c}}^w|=y)$ and $Pr(|e_{\Bar{c}}^w|=z)$ respectively}
%\tcp*[r]{$P_c$ contains all probabilities in $e_c^w$, $P_{c,\Bar{c}}$ and  $P_{\Bar{c}}$ are sets which collect all possible probabilities in $e_{c,\Bar{c}}^w$ and $e_{\Bar{c}}^w$ respectively}
Partition edges into $e_c$, $e_{c,\Bar{c}} $ and $e_{\Bar{c}}$;\\
%\tcp{Calculate approximation of probability of each partition}\\
\For{$x \leftarrow 0$ \KwTo $T_x$}{
%$p(x)\leftarrow $ $Pr(x)$ (Eq.~\ref{appro_each});\\
Store $Pr(|e_c^w|=x)$ in $P_c[x]$;\\
}
\For{$y \leftarrow 0$ \KwTo $T_y$}{
%$p(y)\leftarrow $ $Pr(y)$ (Eq.~\ref{appro_each});\\
Store $Pr(|e_{c,\Bar{c}}^w|=y)$ in $P_{c,\Bar{c}}[y]$;\\
}
\For{$z \leftarrow 0$ \KwTo $T_z$}{
%$p(z)\leftarrow $ $Pr(z)$ (Eq.~\ref{appro_each});\\
Store  $Pr(|e_{\Bar{c}}^w|=z)$ in $P_{\Bar{c}}[z]$;}
%\tcp{Loop of all partitions }
%\For{$d_i \leftarrow 0$ \KwTo 
%$|\mathcal{C}|T_xT_yT_z$}{
\For{$x \leftarrow 0$ \KwTo $T_x$}{
\For{$y \leftarrow 0$ \KwTo $T_y$}{
\For{$z \leftarrow 0$ \KwTo $T_z$}{
$Q_{c}^{xyz} \leftarrow  \frac{x}{(x+y+z)}-(\frac{2x+y}{2(x+y+z)})^2 $ (Eq.~\ref{Qc_xyz});\\
$Sum=Sum +Q_{c}^{xyz}P_c[x]P_{c,\Bar{c}}[y] P_{\Bar{c}}[z]$;\\
}
}
}
%\For{$k_{bet}\leftarrow 0$ \KwTo $|e_{c,\Bar{c}}|$}{

%\For{$k_{out}\leftarrow 0$ \KwTo $|e_{\Bar{c}}|$}{

%$Q_{c}^{xyz} \leftarrow $ modularity %(Eq.~\ref{Qc_xyz});\\
%Calculate probability $Pr(xyz)$ of $Q_{c}^{xyz}$ (Eq.~\ref{appro_xyz});\\
%Calculate expected modularity 
%$E(\textbf{Q})$ (Eq.~\ref{final});\\
%$Sum=Sum + E(\textbf{Q})$
%}
}
%}
%}
\Return{Sum}
\caption{ \DIFdelbegin \DIFdel{$\mathrm{(A)PWP}$ }\DIFdelend \DIFaddbegin \DIFadd{$\mathrm{(F)PWP}$ }\DIFaddend %\mm{(1) Possibilities (2) It is not explained anywhere what $P_c$ etc. are (I think), and what it means to initialize them (3) Space before comments (4) Restore should be just Store? (5) Line 11: I think we should not iterate over $d_i$, but have three nested loops 0 to $T_x$, 0 to $T_y$, and 0 to $T_z$. Because inside the loop we do not use $d_i$, but $c, x, y, z$. (6) I think line 14 is wrong, because Eq. 14 is the whole algorithm. Shouldn't it be, that after lines 12 and 13 we just do $Sum=Sum + Q_{c}^{xyz} Pr(x) Pr(y) Pr(z)$?}
}\label{apwp}
\end{algorithm}\DecMargin{1em}

%____________________
%\textbf{\rev{Computational complexity:}} 
%DIF < the worst-case time complexity of $\mathrm{APWP}$ is polynomial, and depends on two parts. One part is the loop time over all partitions, whose time complexity is $O(|E|^3)$. In the worst case, the number of edges in $e_c$, $e_{c,\Bar{c}}$ and $e_{\Bar{c}}$ are similar to each other. The second part is the approximation time for the  probability of each partition, whose time complexity is $O(|E|^2)$. In the worst case, the number of edges in one of $e_c$, $e_{c,\Bar{c}}$ and $e_{\Bar{c}}$ is much larger than in the other two. We only calculate $pr(x) Pr(y) and Pr(z)$ once for each partitions.
%DIF > the worst-case time complexity of $\mathrm{FPWP}$ is polynomial, and depends on two parts. One part is the loop time over all partitions, whose time complexity is $O(|E|^3)$. In the worst case, the number of edges in $e_c$, $e_{c,\Bar{c}}$ and $e_{\Bar{c}}$ are similar to each other. The second part is the approximation time for the  probability of each partition, whose time complexity is $O(|E|^2)$. In the worst case, the number of edges in one of $e_c$, $e_{c,\Bar{c}}$ and $e_{\Bar{c}}$ is much larger than in the other two. We only calculate $pr(x) Pr(y) and Pr(z)$ once for each partitions.
The worst-case time complexity of \DIFdelbegin \DIFdel{$\mathrm{APWP}$ }\DIFdelend \DIFaddbegin \DIFadd{$\mathrm{FPWP}$ }\DIFaddend is $O(km^3)$. We can break \rev{the analysis} down into two main \rev{parts.}
Lines 3 to 10 concern the probabilities for each partition. The time complexities of $Pr(x)$, $Pr(y)$, and $Pr(z)$ are $O(m^2)$. From Equation~\ref{exp_mod_partitioned}, we can see that $x$ ranges from 1 to $T_x$, $y$ ranges from 1 to $T_y$, and $z$ ranges from 1 to $T_z$. $T_x$, $T_y$, and $T_z$ sum to $m$, so the overall complexity is $O(m^3)$. 
Lines 11 to 15 involve looping over all partitions, which has a time complexity of $O(m^3)$. Notice that in the worst-case scenario, the number of edges in $e_c$, $e_{c,\Bar{c}}$, and $e_{\Bar{c}}$ are similar to each other. Here the number of edges in $e_c$, $e_{c,\Bar{c}}$, and $e_{\Bar{c}}$ have the same upper bound $m$.

Both parts are inside the community loop on line 2, so the time complexity of the whole algorithm is $O(km^3)$, where  $k$ is the number of communities. Notice that the time complexity solely depends on the \rev{number} of communities and edges, and remains unaffected by the assignment of edge probabilities\rev{.}

\section*{Evaluation}
\label{sec:eval}

The brute-force method produces the correct result and can thus be used to evaluate the other approaches. However, it can only be used on very small networks. Therefore, we start our evaluation showing that our method is as accurate as the brute-force method,
%a full computation of expected modularity, 
but is several orders of magnitude faster.  
In this way, we can then use our method to evaluate the other possible ways to estimate expected modularity: weighting, thresholding, and sampling. We also study the behavior of our algorithm when we vary the number of communities, the distribution of community sizes, and the type of input network.

All the experiments except the comparison with the brute-force method have been performed on a macOS system, with %32G memory and 
max CPU frequency 2.4GHz. %he machine that runs 
The comparison with the brute-force method has been performed on a Linux system, with 
%31G memory and 
max CPU frequency 5GHz. Both systems have the same 32GB memory capacity.
%All the code is in Python and available at \href{https://github.com/XINS3/Expected-modularity-calculation-over-uncertain-graph}{https://github.com/XINS3/Expected-modularity-calculation-over-uncertain-graph}.% \mm{add the link to the repository}.

\emph{\rev{Data.}} In our experiments we use both synthetic datasets, to control and examine the properties of the data that may affect running time and accuracy of the tested algorithms, and real-world datasets. %, to demonstrate the practical use of our methods.

To generate synthetic datasets, we use Stochastic Block Model (SBM), Forest Fire Network (FFN), Barabási-Abert (BA), Small World (SW), and Erdős-Rényi (ER). The parameters used to generate specific random networks are specified later in this section for each experiment.

As real datasets, we use the Enron email network, %a Facebook network
\DIFaddbegin \DIFadd{an Online Social Network (OSN), }\DIFaddend a \DIFaddbegin \DIFadd{P}\DIFaddend rotein-\DIFaddbegin \DIFadd{P}\DIFaddend rotein \DIFaddbegin \DIFadd{I}\DIFaddend nteraction (PPI) network, and a \DIFaddbegin \DIFadd{C}\DIFaddend ollaboration network, summarized in Table \ref{fig:datasets}. The data sources used to generate these networks have been commonly used in the literature on probabilistic networks. In particular, two datasets (Enron, PPI) have been directly taken from the literature, while for those not provided by the authors we obtained probabilistic networks from the same sources following similar procedures and reproducing similar network statistics. For \DIFaddbegin \DIFadd{OSN}\DIFaddend , Collaboration, and PPI data
%, to keep the original topology structure, 
we included all the nodes in selected clusters detected by modularity optimization and all edges between those nodes. In our experiments, these clusters are also used as input to compute expected modularity. 

\begin{enumerate}
\item 
\noindent\textbf{Enron:}  this dataset consists of emails sent between employees of Enron between 1999 and 2001. Nodes represent employees and there is an edge between two nodes if at least one email has been exchanged between them. The original dataset with edge probabilities is provided by the authors \cite{kaveh2021defining}.
\item 
\noindent\textbf{Protein-\DIFaddbegin \DIFadd{P}\DIFaddend rotein \DIFaddbegin \DIFadd{I}\DIFaddend nteraction (PPI):} the dataset contains nodes that represent proteins, edges that represent the interactions between two proteins, and associated probabilities. This dataset is extracted from a protein database that directly provides interaction probabilities
\cite{krogan2006global}. 
\item  
\noindent\textbf{Online \DIFaddbegin \DIFadd{Social Network (OSN)}\DIFaddend :} the dataset is obtained from a weighted Facebook-like \DIFaddbegin \DIFadd{social }\DIFaddend network, that originates from an online community for students at the University of California, Irvine. The data, downloaded from toreopsahl.com, includes the users who sent or received at least one message.  Probabilities are computed applying an exponential cumulative distribution function (CDF) of mean 2 to the weights \cite{li2015recursive}.
We sample nodes and edges from the original networks. 
\item 
\noindent\textbf{Collaboration:} the dataset is obtained from  a weighted network of coauthorships between scientists posting preprints of the astrophysics archive at www.arxiv.org from 1995-1999
%\footnote{https://websites.umich.edu/$\sim$mejn/netdata}
\cite{newman2001structure}. We computed the probabilities using the same method as for the \DIFaddbegin \DIFadd{OSN }\DIFaddend  data. 

%\item \rev{Lesmis dataset: the dataset is a weighted network of appearances of characters in Victor Hugo's novel "Les Miserables".  Nodes represent characters as indicated by the labels and edges connect any pair of characters that appear in the same chapter of the book.  The values on the edges are the number of such appearances. We download this dataset from (https://websites.umich.edu/$\sim$ mejn/netdata/)
%}
\end{enumerate}

%\rev{%DIF > we use the same sampling method in \cite{kaveh2021defining}. To provide a community assignment for the computation of modularity, we use a modularity optimization on the original deterministic networks, where the status of edges are only exist and not exist. 
%Table \ref{fig:datasets} summarizes the real datasets.
%}

\begin{table}[ht]
   \centering

  %DIF > \setlength{\tabcolsep}{0pt} % let "tabular*" figure out intercolumn widths
    \begin{tabular}{ccrr}
    \toprule
     \rev{Name}& \rev{Nodes}&\rev{Edges}&\rev{Clusters}\\
    \midrule
    %DIF > 6 & 2&9.5e-16& 0.005086& 0.022277 \\
    %DIF > 6 & 2&1.6e-15& 0.005078& 0.022317  \\

   %DIF > 16 & 4&3.4e-13& 0.235500& 92.130696  \\
   \rev{Enron}& \rev{524 }&\rev{833 }&\rev{3}\\

   %DIF > 19 & 0.5560&0.5560& 0.5560 \\
   %DIF > 20 & 5&7.6e-13&2.784840&2356.276487 \\
    %DIF > 20 & 6&2.8e-13&2.720410&2349.818362 \\
    %DIF > 17 &0.3634&0.3634&0.3634\\
    \rev{PPI}&\rev{593}&\rev{1185}&\rev{5}\\
    %\rev{Lesmis}&\rev{77}&\rev{254}&\rev{5}\\

    %DIF > 3 & 0.1125&0.1125&0.1125  \\
   \DIFaddbegin \DIFadd{OSN }\DIFaddend & \rev{306}&\rev{1217}&\rev{4}\\
   %DIF > 10 & 3&1.1e-15& 0.019749& 0.368506  \\
   \rev{Collaboration}&\rev{523}&\rev{1224}&\rev{2}\\

    \bottomrule
    \end{tabular}
    \caption{Summary of real datasets.} 
    \label{fig:datasets}
    \end{table}

\subsection*{Accuracy and execution time}

%\subsection*{Comparison with brute-force method}
To 
%show 
\DIFaddbegin \DIFadd{verify }\DIFaddend 
the accuracy of our method, we compare the values of expected modularity computed using $\mathrm{PWP}$ and \DIFdelbegin \DIFdel{$\mathrm{APWP}$ }\DIFdelend \DIFaddbegin \DIFadd{$\mathrm{FPWP}$ }\DIFaddend with the correct results, that we obtain using a brute-force approach iterating through all the possible worlds. It is worth noting that \DIFaddbegin \DIFadd{both $\mathrm{PWP}$ and $\mathrm{FPWP}$ are exact algorithms, that is, they are }\DIFaddend expected to compute the same result as the brute-force method \DIFaddbegin \DIFadd{(modulo numerical approximation)}\DIFaddend: the experimental results are provided to show the correctness of \DIFaddbegin \DIFadd{our }\DIFaddend mathematical derivations.

%on which $\mathrm{PWP}$ is based. \DIFdelbegin \DIFdel{$\mathrm{APWP}$ }\DIFdelend \DIFaddbegin \DIFadd{$\mathrm{FPWP}$ }\DIFaddend computes the expected modularity with the aim of further reducing computing time. Therefore, this experiment also aims to inspect the accuracy of the approximation.\chroh{fix}

%For this experiment, described in Table~\ref{fig:accuracy} and Table~\ref{fig:comparison}, 
For the first experiment,
we generate networks with three communities and varying sizes, using a stochastic block model (also known as planted-community structure network) \cite{condon2001algorithms}, with the parameters shown in Table~\ref{fig:para1}. We adjust $p_{in}$ and $p_{out}$ to control the size of the networks. For the second experiment, we generate random probabilistic networks, using Erdős–Rényi model, with different numbers of edges and we fix the number of communities to 5. % \mm{This may raise questions, because with such small communities (e.g. with 2 nodes) the edge density doesn't really make sense - it will never be .8, only 1 or 0. Also, we call it $p$, so that's the input probability of in-community edges, is not the actual density. And we change the probabilities when we change the size, which would need an explanation. Some values also look random (such as the 0.04 only for 21 edges but 0.03 for all the other cases). My suggestions: (1) remove Table 2 (2) pick fixed probabilities for all sizes, \emph{for example} p$_{in}$ .7 and p$_{out}$ .3 (motivate this high number so that we have some inter-community edges) and just report them in the text, without a separate table (3) remove size 3 from Table 3 (also because we do not have it in Table 4 anyway.)}
%XIN: PLEASE PROVIDE THE PARAMETERS OF THE GENERATION METHOD (p$_{in}$, etc.) . 
Table~\ref{fig:accuracy} shows that 
%DIF < the accuracy of our methods $APWP^{EMOD}$ and $PWP^{EMOD}$. Comparing the column 2 to 4, we can find 
%DIF > the accuracy of our methods $FPWP^{EMOD}$ and $PWP^{EMOD}$. Comparing the column 2 to 4, we can find 
the results calculated by $\mathrm{PWP}$ and \DIFdelbegin \DIFdel{$\mathrm{APWP}$ }\DIFdelend \DIFaddbegin \DIFadd{$\mathrm{FPWP}$ }\DIFaddend are always the same as the true values, calculated by the brute-force method, down to several significant digits (four in the table).
%Due to the computational challenges involved in using the brute-force method for networks with more than a few edges, we discontinue testing the brute-force method. However, w
We also observe that even for larger networks where we can no longer use the brute-force method, 
the results obtained by \DIFdelbegin \DIFdel{$\mathrm{APWP}$ }\DIFdelend \DIFaddbegin \DIFadd{$\mathrm{FPWP}$ }\DIFaddend is 
%practically 
the same as the one produced by $\mathrm{PWP}$\DIFaddbegin \DIFadd{, as expected.}\DIFaddend
\begin{table}[ht]
   \centering

    \begin{tabular}{ccrrrr}
    \toprule
     $m$ & $n$&$k$& $\mathrm{nc}$&$p_{in}$&$p_{out}$  \\
    \midrule
    %6 & 2&9.5e-16& 0.005086& 0.022277 \\
    %6 & 2&1.6e-15& 0.005078& 0.022317  \\
    %3 &6& 3&2&0.8&0.03  \\
   9 & 9&3&3&0.8&0.03  \\
   %10 & 3&1.1e-15& 0.019749& 0.368506  \\

   %16 & 4&3.4e-13& 0.235500& 92.130696  \\
   14 & 12&3&4&0.8&0.03 \\
   %19 & 0.5560&0.5560& 0.5560 \\
   %20 & 5&7.6e-13&2.784840&2356.276487 \\
    %20 & 6&2.8e-13&2.720410&2349.818362 \\
    %17 &15&3&5&0.5&0.03\\
    21&15&3&5&0.6&0.04\\
    25 & 18&3&6&0.5&0.03\\
    %29 & 18 & 3& 6& 0.6& 0.03\\
    %31 & 18 & 3 & 6& 0.7&0.02\\
    35 & 21 & 3 &7 & 0.5 & 0.03\\
%    49 & 24 & 3 & 8 & 0.6 &0.03\\
    \bottomrule
    \end{tabular}
    \medskip

    \begin{tabular}{ll}

    $m$ & number of edges\\ 
    $n$ & number of nodes \\
    $k$ & number of communities \\
    $\mathrm{nc}$ & number of nodes in each community \\
    $p_{in}$ & edge density within communities\\
    $p_{out}$ & edge density between communities\\

    \end{tabular}

     \caption{Experimental parameters.}  
    \label{fig:para1}
    \end{table}

\begin{table}[ht]
\captionsetup{justification=centering}
   \centering

    \begin{tabular}{ccrr}
    \toprule
     $m$ & \DIFaddFL{$Q^{\mathrm{PWP}}$}\DIFaddendFL &\DIFdelbeginFL \DIFdelFL{$Q^{\mathrm{APWP}}$}\DIFdelendFL \DIFaddbeginFL \DIFaddFL{$Q^{\mathrm{FPWP}}$}\DIFaddendFL & \DIFaddbeginFL \DIFaddFL{$Q^{BF}$}\DIFaddendFL  \\
    \midrule
    %6 & 2&9.5e-16& 0.005086& 0.022277 \\
    %6 & 2&1.6e-15& 0.005078& 0.022317  \\
    %3 & 0.1125&0.1125&0.1125  \\
   9 & 0.3900&0.3900&0.3900 \\
   %10 & 3&1.1e-15& 0.019749& 0.368506  \\

   %16 & 4&3.4e-13& 0.235500& 92.130696  \\
   14 & 0.4401 &0.4401 &0.4401 \\
   %19 & 0.5560&0.5560& 0.5560 \\
   %20 & 5&7.6e-13&2.784840&2356.276487 \\
    %20 & 6&2.8e-13&2.720410&2349.818362 \\
    %17 &0.3634&0.3634&0.3634\\
    21&0.4257&0.4257&0.4257\\
    25 & 0.4455&0.4455&0.4455\\
    %29& 0.4299 & 0.4299 & --\\
    %31&  0.4809&0.4809&--\\
    35&  0.4720&0.4720&--\\
    \bottomrule
    \end{tabular}

\medskip

    \begin{tabular}{ll}

    $m$ & number of edges\\ 
    \DIFaddbeginFL \DIFaddFL{$Q^{\mathrm{PWP}}$}\DIFaddend  & expected modularity score \DIFaddbegin \DIFadd{computed by }\DIFaddend $\mathrm{PWP}$ \\
    \DIFdelbeginFL \DIFdelFL{$S^{\mathrm{APWP}}$ }\DIFdelendFL \DIFaddbeginFL \DIFaddFL{$Q^{\mathrm{FPWP}}$ }\DIFaddendFL & expected modularity score \DIFaddbegin \DIFadd{computed by }\DIFaddend \DIFdelbeginFL \DIFdelFL{$\mathrm{APWP}$ }\DIFdelendFL \DIFaddbeginFL \DIFaddFL{$\mathrm{FPWP}$ }\DIFaddendFL \\
    \DIFaddbeginFL \DIFaddFL{$Q^{BF}$}\DIFaddend & expected modularity score \DIFaddbegin \DIFadd{computed by the }\DIFaddend brute-force method\\

    \end{tabular}

     \caption{Expected modularity computed using the brute-force method, $\mathrm{PWP}$, and \DIFdelbeginFL \DIFdelFL{$\mathrm{APWP}$}\DIFdelendFL \DIFaddbeginFL \DIFaddFL{$\mathrm{FPWP}$}\DIFaddendFL
     %, only on small networks because of the time complexity of the brute-force method
     .}
    \label{fig:accuracy}

      \end{table}

\begin{figure}[ht]
    \centering
    \includegraphics[width=.5\textwidth]{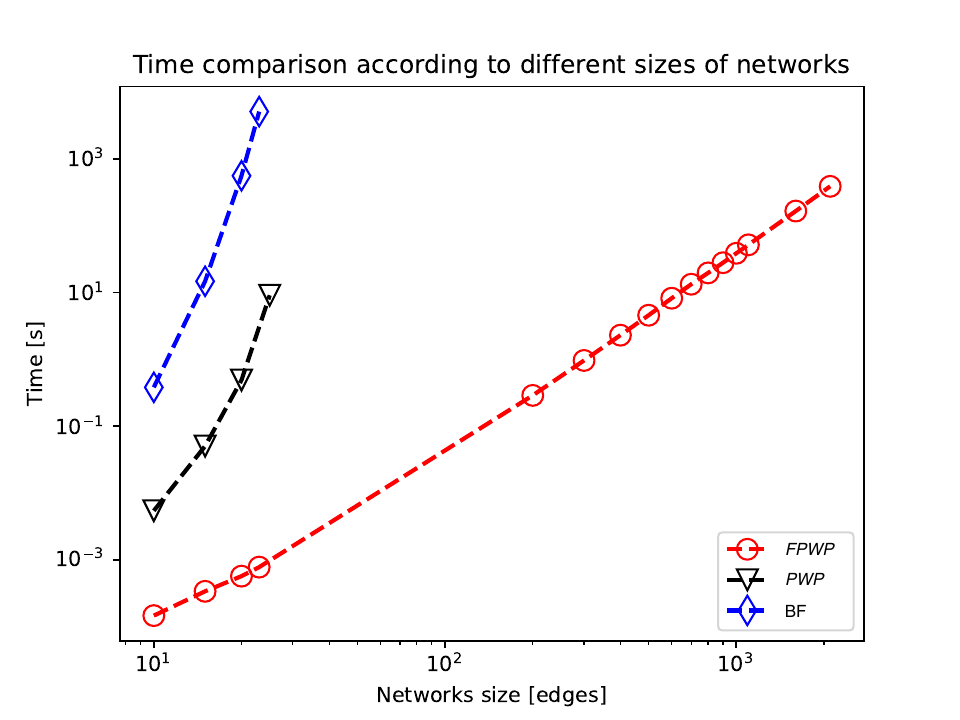}
    \caption{Running time of brute-force, $\mathrm{PWP}$, and \DIFdelbeginFL \DIFdelFL{$\mathrm{APWP}$}\DIFdelendFL \DIFaddbeginFL \DIFaddFL{$\mathrm{FPWP}$}\DIFaddendFL , with log axes. %\mm{Make sure these correspond to the values in the final Table 4. And in this and all the other plots please use different colors but also different line styles for the different curves (continuous, dashed, dotted, \dots), so that the paper is also readable in black and white (and in case by color blind people).}
    }
    \label{fig:size_net}
\end{figure}

Table~\ref{fig:comparison} and Figure~\ref{fig:size_net} show that both our methods are multiple orders of magnitude faster than the brute-force approach. 
%In particular, Table~\ref{fig:comparison} shows the execution times for the same networks used in Figure~\ref{fig:accuracy} where all three methods can be run, and 
Figure~\ref{fig:size_net} adds larger networks to test the behavior of \DIFdelbegin \DIFdel{$\mathrm{APWP}$ }\DIFdelend \DIFaddbegin \DIFadd{$\mathrm{FPWP}$ }\DIFaddend where the two other methods cannot be used. %DIF < The lowest running time values are highlighted in bold in the last three columns, indicating that $APWP^{EMOD}$ generally exhibits the shortest running time.
%DIF > The lowest running time values are highlighted in bold in the last three columns, indicating that $FPWP^{EMOD}$ generally exhibits the shortest running time.
Both $\mathrm{PWP}$ and the brute-force method have an exponential time complexity, although, $\mathrm{PWP}$ is significantly faster.  Figure~\ref{fig:size_net} confirms the polynomial time complexity of \DIFdelbegin \DIFdel{$\mathrm{APWP}$}\DIFdelend \DIFaddbegin \DIFadd{$\mathrm{FPWP}$}\DIFaddend .

% we can see that although the running time of both 

%DIF < In this experiment, we compare our two methods $APWP^{EMOD}$ and $PWP^{EMOD}$ with the brute-force method (BF) to show our methods' (especially approximation $APWP^{EMOD}$'s) time efficiency.
%DIF > In this experiment, we compare our two methods $FPWP^{EMOD}$ and $PWP^{EMOD}$ with the brute-force method (BF) to show our methods' (especially approximation $FPWP^{EMOD}$'s) time efficiency.

\begin{table}[ht]
\captionsetup{justification=centering}
   \centering

    \begin{tabular}{crrr}
    \toprule
     $m$ & $T^{PWP}$&\DIFdelbeginFL \DIFdelFL{$T^{APWP}$}\DIFdelendFL \DIFaddbeginFL \DIFaddFL{$T^{FPWP}$}\DIFaddendFL & $T^{BF}$ \\
    \midrule
    %6 & 2&9.5e-16& 0.005086& 0.022277 \\
    %6 & 2&1.6e-15& 0.005078& 0.022317  \\
%    3 & \textbf{0.0003} & 0.0004 &0.0453  \\ I CAN EXPLAIN
   9 &0.0020 &0.0001& 0.1201\\
   %10 & 3&1.1e-15& 0.019749& 0.368506  \\

   %16 & 4&3.4e-13& 0.235500& 92.130696  \\
   14  &0.0082&0.0004 &8.7528\\
   %19 & 0.0591&\textbf{0.0016}& 555.6865\\
   %20 & 5&7.6e-13&2.784840&2356.276487 \\
    %20 & 6&2.8e-13&2.720410&2349.818362 \\
    %17 &0.8258 &\textbf{0.0009}& 104.9380\\
    21 &0.1916 &0.0004& 2401.2359\\
    25 & 3.0294&0.0006&59956.8956\\
%    29 (renew) & 73.7829 & \textbf{0.0231}& --\\
  %  31 (renew) & 2804.5993 & \textbf{0.0101}&--\\
    35 & 226.5636& 0.0030&--\\
%    49 & -- & \textbf{0.0260}&--\\
    \bottomrule
    \end{tabular}

\medskip

    \begin{tabular}{ll}

    $m$ & number of edges\\ 

    $T^{PWP}$ & running time of $\mathrm{PWP}$ \\
    \DIFdelbeginFL \DIFdelFL{$T^{APWP}$ }\DIFdelendFL \DIFaddbeginFL \DIFaddFL{$T^{FPWP}$ }\DIFaddendFL & running time of \DIFdelbeginFL \DIFdelFL{$\mathrm{APWP}$ }\DIFdelendFL \DIFaddbeginFL \DIFaddFL{$\mathrm{FPWP}$ }\DIFaddendFL \\
    $T^{BF}$ & running time of brute-force method\\
%    $t$ & computational complexity for approximation [s]\\ 
%    $t^*$ & computational complexity for exact value [s]
    \end{tabular}

     \caption{\DIFdelbeginFL \DIFdelFL{\rev{Running time of brute-force%\mm{It is sometimes called \emph{brute force}, sometimes \emph{brute-force}. Check these. I think \emph{brute-force} is an adjective, so it can be \emph{brute-force method}.}
, $\mathrm{PWP}$, and $\mathrm{APWP}$.}  }\DIFdelendFL \rev{Running time of brute-force%\mm{It is sometimes called \emph{brute force}, sometimes \emph{brute-force}. Check these. I think \emph{brute-force} is an adjective, so it can be \emph{brute-force method}.}
, $\mathrm{PWP}$, and $\mathrm{FPWP}$.}  }
    \label{fig:comparison}

      \end{table}

In the next sections we only use \DIFdelbegin \DIFdel{$\mathrm{APWP}$}\DIFdelend \DIFaddbegin \DIFadd{$\mathrm{FPWP}$}\DIFaddend , given that  it is both accurate and much faster than the brute-force method and $\mathrm{PWP}$.

\subsection*{Comparison with alternative methods}

In this section we show that both weighting (that is, treating the probabilities as weights) and thresholding (that is, considering the network as a deterministic one after keeping only high-probability edges) lead to wrong estimations of expected modularity. On the contrary, sampling can be used to get more and more accurate results by increasing execution time, so we also study how the balance between time and accuracy when using a sampling approach compares to our algorithm.

\subsubsection*{Weighting}

\begin{figure}[ht]
\captionsetup{justification=centering}
    \centering
    \includegraphics[width=.5\textwidth]{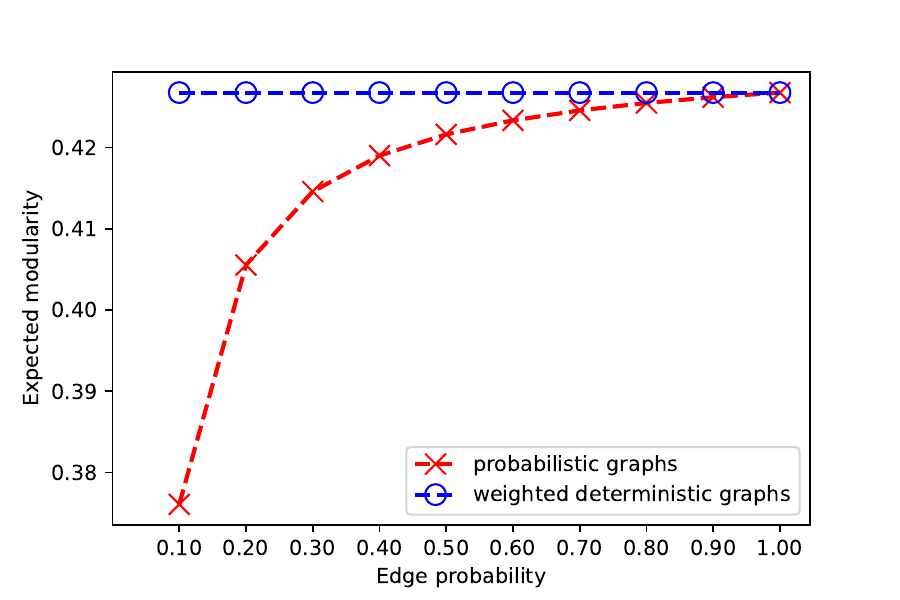}
    \caption{Weighted and expected modularity calculated on the same networks.}
    \label{fig:weight}
\end{figure}

In this experiment, we show that directly regarding probabilistic networks as weighted deterministic networks leads to a wrong expected modularity calculation. In particular, we first generate a base network using stochastic block modeling, where the parameters are $k=3$, $\mathrm{nc}=9$, $p_{in}=0.72$ and $p_{out}=0.12$%\mm{These in practice result into separate cliques. We should do this with lower $p_{in}$ and higher $p_{out}$.}
.  From this base network, we generate multiple probabilistic networks by assigning probabilities to its edges. \rev{I}n each probabilistic network we assign the same probability to all the edges: for example, in Figure~\ref{fig:weight}, the network corresponding to the value 0.20 on the $x$ axis has all its edges having probability 0.20.

Figure~\ref{fig:weight}  shows expected modularity computed using our algorithm (red) and the result obtained using weighted modularity. We can see that if we directly regard probabilistic networks as weighted deterministic networks, the final results are always the same for the different tested networks, and also different from expected modularity except for the case when the network is deterministic (where all edge probabilities equal 1). 
%And if we consider the modularity formula (Eq~\ref{e1}), we find that the weight will be offset if all edge weights are the same.
%Considering a network has the same edge probabilities, if we regard it as weighted deterministic network, we will take weight into the  number of edges, node degrees and adjacency matrix, then the weight will be offset in modularity equation (Eq.~\ref{e1}), and finally the network with same edge probabilities will have the same modularity of unweighted network. However, if we consider it as probabilistic network and calculated the expected modularity, such edge probabilities will not be offset.

\subsubsection*{Thresholding}
In this experiment, we compare expected modularity (as computed by \DIFdelbegin \DIFdel{$\mathrm{APWP}$}\DIFdelend \DIFaddbegin \DIFadd{$\mathrm{FPWP}$}\DIFaddend ) with the modularity obtained after thresholding, showing that the choice of threshold largely influences the final results, which are also not accurate in general.

\begin{figure}[ht]
    \centering
     \includegraphics[width=.5\textwidth]{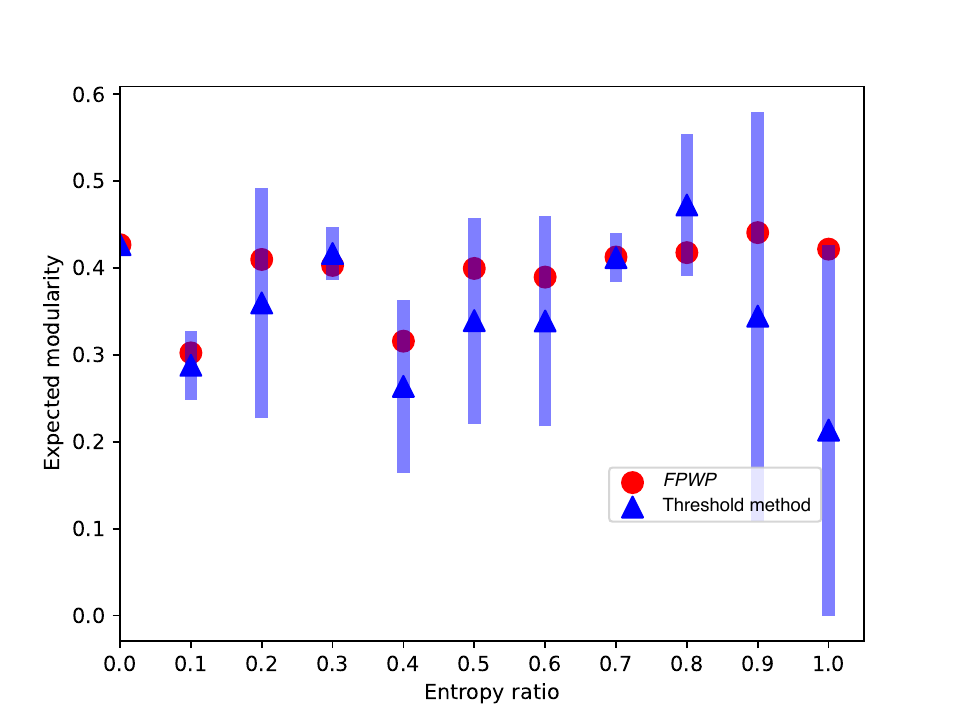}
     \caption{Mean and expected modularity calculated by \DIFdelbeginFL \DIFdelFL{$\mathrm{APWP}$ }\DIFdelendFL \DIFaddbeginFL \DIFaddFL{$\mathrm{FPWP}$ }\DIFaddendFL and using the threshold method.}
    \label{fig:thre}
\end{figure}

 \begin{figure}[ht]
\centering

\begin{subfigure}{.3\textwidth}
    \includegraphics[width=\textwidth]{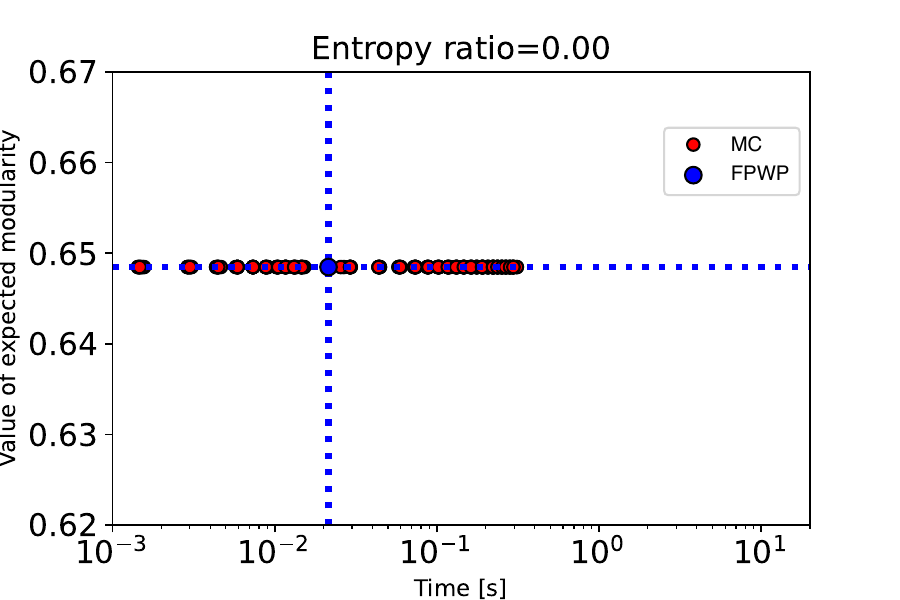}
      \caption
            {\emph{CCS network}}    
            \label{fig:n11}
 \end{subfigure}
\begin{subfigure}{.3\textwidth}
    \includegraphics[width=\textwidth]{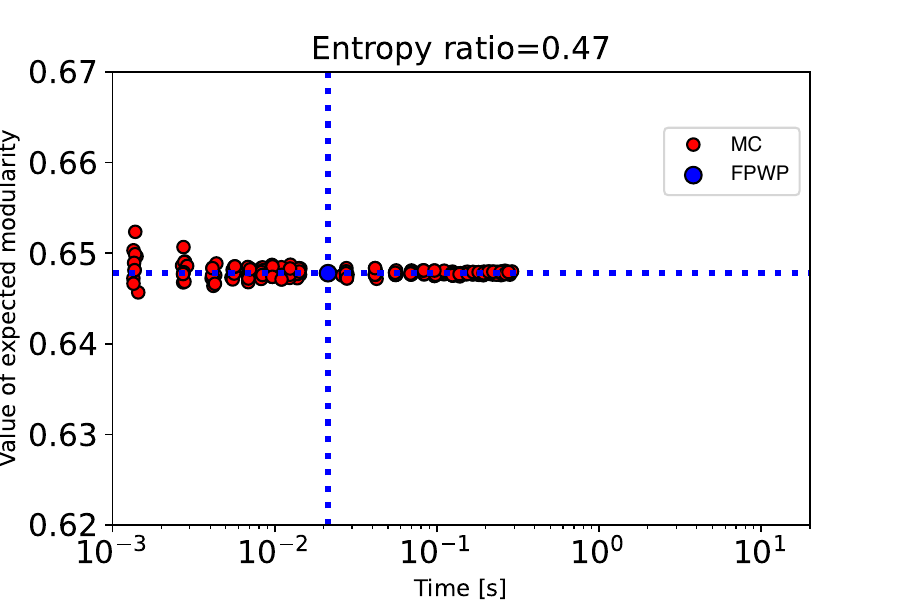}
      \caption
            {\emph{CCS network}}   
            \label{fig:n9}
 \end{subfigure}
 \begin{subfigure}{.3\textwidth}
    \includegraphics[width=\textwidth]{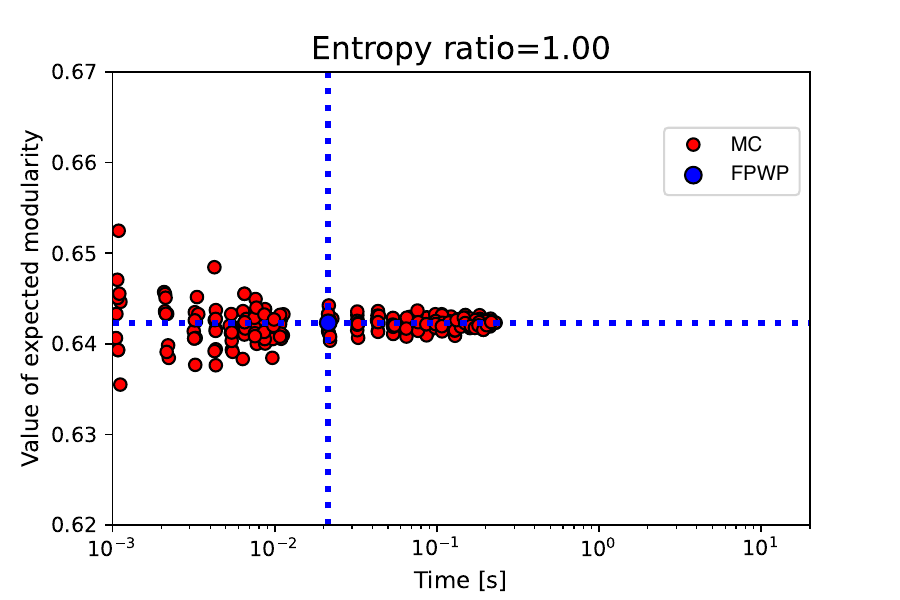}
     \caption
            {\emph{CCS network}}   
            \label{fig:n1}
 \end{subfigure}
\begin{subfigure}{.3\textwidth}
    \includegraphics[width=\textwidth]{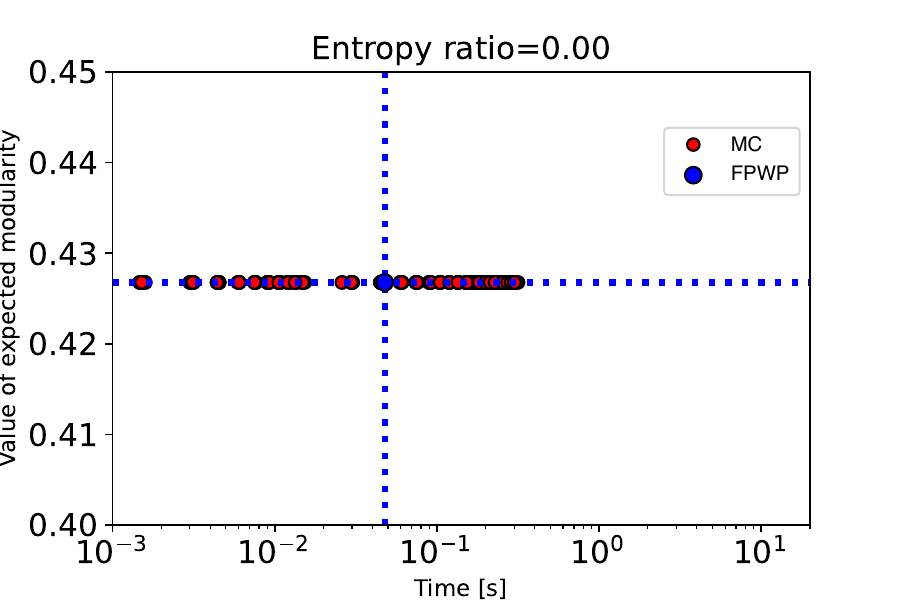}
     \caption
            {\emph{LCCS network}}    
            \label{fig:n12}
 \end{subfigure}
\begin{subfigure}{.3\textwidth}
    \includegraphics[width=\textwidth]{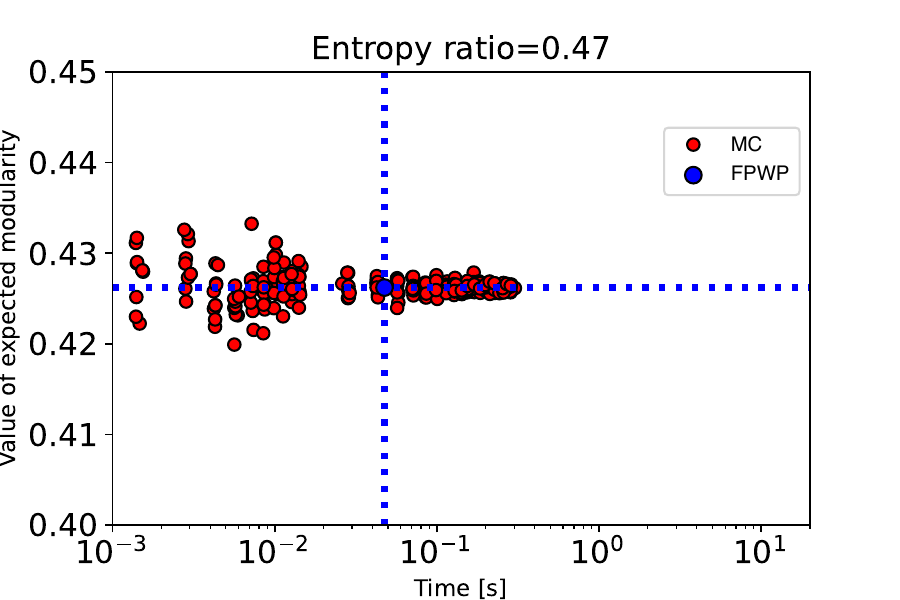}
     \caption
            {\emph{LCCS network}}    
            \label{fig:n10}
 \end{subfigure}
\begin{subfigure}{.3\textwidth}
    \includegraphics[width=\textwidth]{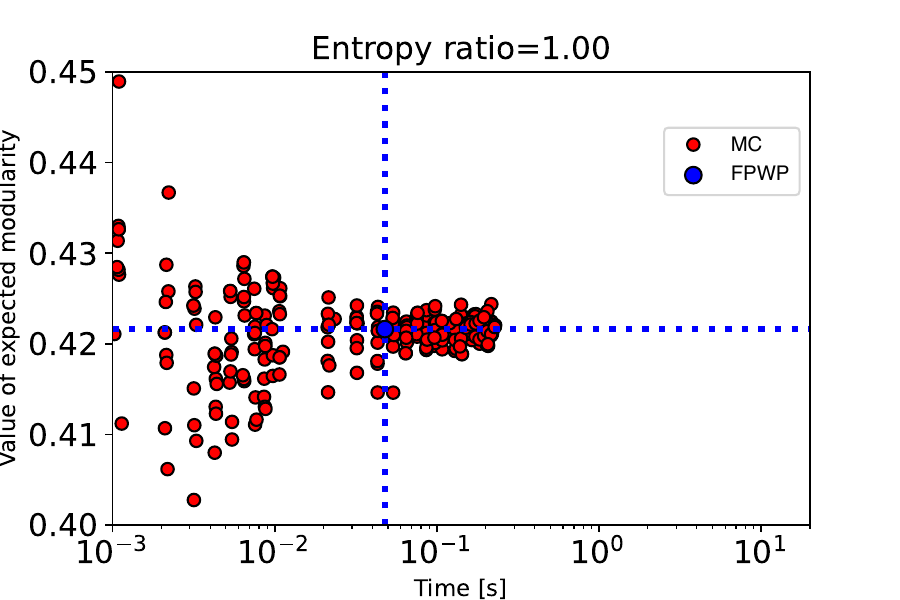}
     \caption
            {\emph{LCCS network}}    
            \label{fig:n2}
 \end{subfigure}
\caption{\rev{Running time and expected modularity value calculated using sampling and $\mathrm{FPWP}$ for networks with a more clear (first row) and less clear (second row) modular structure, with different  entropies.}}
\label{fig:scatter}

 \end{figure}

Here we generate planted community structure networks with 110 edges and 3 communities. All networks have the same topology, generated by a stochastic block model with %$k=3$, $l=9$, 
$p_{in}=0.72$ and $p_{out}=0.12$. We then generate different probabilistic networks by assigning random probabilities to the edges so that, for different networks, entropy ratio ranges from 0 to 1, and randomly assign those probabilities to the edges. For each network, we compute the modularities obtained using different thresholds (from 0.1 to 1). In Figure~\ref{fig:thre} we plot their mean (blue circles) and standard deviation (thick vertical lines). We can see that except for networks with low entropy ratios (that is, networks very close to deterministic networks), the modularities obtained through thresholding are %both 
far from the correct values (red line, calculated using \DIFdelbegin \DIFdel{$\mathrm{APWP}$}\DIFdelend \DIFaddbegin \DIFadd{$\mathrm{FPWP}$}\DIFaddend ). % and largely dependent on the chosen threshold.
%FIONA: cannot see that it is dependent on threshold from Fig 5 (though clearly the thresholding technique is dependent on the threshold chosen), so commented out a part of sentence.
%DIF < , which shows that $APWP^{EMOD}$ has higher accuracy.
%DIF > , which shows that $FPWP^{EMOD}$ has higher accuracy.

\subsubsection*{Sampling} \label{sampling_sec}

Thanks to the fact that using \DIFdelbegin \DIFdel{$\mathrm{APWP}$ }\DIFdelend \DIFaddbegin \DIFadd{$\mathrm{FPWP}$ }\DIFaddend we can compute expected modularity in polynomial time for a given input network, we can now test 
the ability of the sampling method to accurately estimate expected modularity, and study the factors influencing the balance between accuracy and execution time when sampling is used.

%how 
%DIF < accurate the result obtained using sampling is when we use the same time taken by $\mathrm{APWP}$.  We also %test how much longer we have to wait for the
%DIF > accurate the result obtained using sampling is when we use the same time taken by $\mathrm{FPWP}$.  We also %test how much longer we have to wait for the
%Monte Carlo (MC) to have  a high chance of getting a result close to the correct value of expected modularity.} 
%

We use two kinds of random networks. One is characterized by a pronounced community structure, featuring dense intra-community edges and sparse inter-community edges (\emph{CCS network}). The other exhibits a less distinct community structure, with less intra-community edges and more inter-community edges (\emph{LCCS network}). 
Both networks contain 27 nodes and 3 communities.
%with similar number of edges by moving a few edges from within a community to between community from \emph{CCS network} to \emph{LCCS network}. 
The \emph{CCS network} and \emph{LCCS network} are constructed with stochastic block models where the parameters are, respectively, $k=3$, $\mathrm{nc}=9$, $p_{in}=0.99$, and $p_{out}=0.01$, and $k=3$, $\mathrm{nc}=9$, $p_{in}=0.72$ and $p_{out}=0.12$. Notice that the CSS network closely approximates three cliques. When the entropy ratio equals 1.00, all edge probabilities are set to 0.50; when the entropy ratio equals 0.47, all edge probabilities are set to 0.90; when the entropy ratio is 0.00, all edges are deterministic. %And we control the length of y-axis fixed in both two networks.

Figure~\ref{fig:scatter}  shows the execution time (vertical blue line) and the computed expected modularity (blue horizontal line) for our algorithm. Each red circle is the result of a different execution of the sampling method. 
%The red nodes aligned in 'bands' are the result of a different execution of the sampling method with the same input $\theta$ in Algorithm~\ref{sampling}. 
Specifically, the red nodes arranged in \DIFaddbegin \DIFadd{`}\DIFaddend bands' are the result of different executions of the sampling method, all employing the same input $\theta$ as defined in Algorithm~\ref{sampling}.
For different executions, we have let the sampling method run for specific amounts of time (starting from the time needed by our algorithm), so that we could inspect the balance between execution time and likelihood to return an accurate result for the sampling method. For example, in Figure~\ref{fig:n2} we can see eight executions of the sampling method close to the blue vertical line: these were stopped after the same time used by our algorithm to return, and their $y$ value is the expected modularity they returned.  
From the figure we can see that, for different edge probability entropies, sampling always converges to the result produced by \DIFdelbegin \DIFdel{$\mathrm{APWP}$}\DIFdelend \DIFaddbegin \DIFadd{$\mathrm{FPWP}$}\DIFaddend .
%, which also indicates the accuracy of our method. THIS, WE HAVE ALREADY SHOWN, NO NEED TO RE-PROBLEMATISE IT
Except for the case with no uncertainty (entropy ratio=0), sampling takes longer than \DIFdelbegin \DIFdel{$\mathrm{APWP}$ }\DIFdelend \DIFaddbegin \DIFadd{$\mathrm{FPWP}$ }\DIFaddend to converge toward the right results.
%, which can show our method's time efficiency. 
By looking at Figures~\ref{fig:n11},~\ref{fig:n9},~\ref{fig:n1} or Figures~\ref{fig:n12},~\ref{fig:n10},~\ref{fig:n2}, we find that when  entropy ratio increases (larger uncertainty), sampling converges more slowly. If we compare two figures on the same \rev{column}, e.g. \ref{fig:n9} and \ref{fig:n10}, we find that when the community structure is less clear, the convergence time of sampling is longer.

\subsection*{Factors influencing execution time of \DIFdelbegin \DIFdel{$\mathrm{APWP}$}\DIFdelend \DIFaddbegin \DIFadd{$\mathrm{FPWP}$}\DIFaddend }
\begin{figure}[ht]
    \centering
    %\includegraphics[scale=0.5]{figure/our_method_N_clusters_0503.png}
     %DIFDELCMD < \includegraphics[scale=0.5]{figure/T_cluster_num.pdf}
%DIFDELCMD <     %%%
  \includegraphics[width=.5\textwidth]{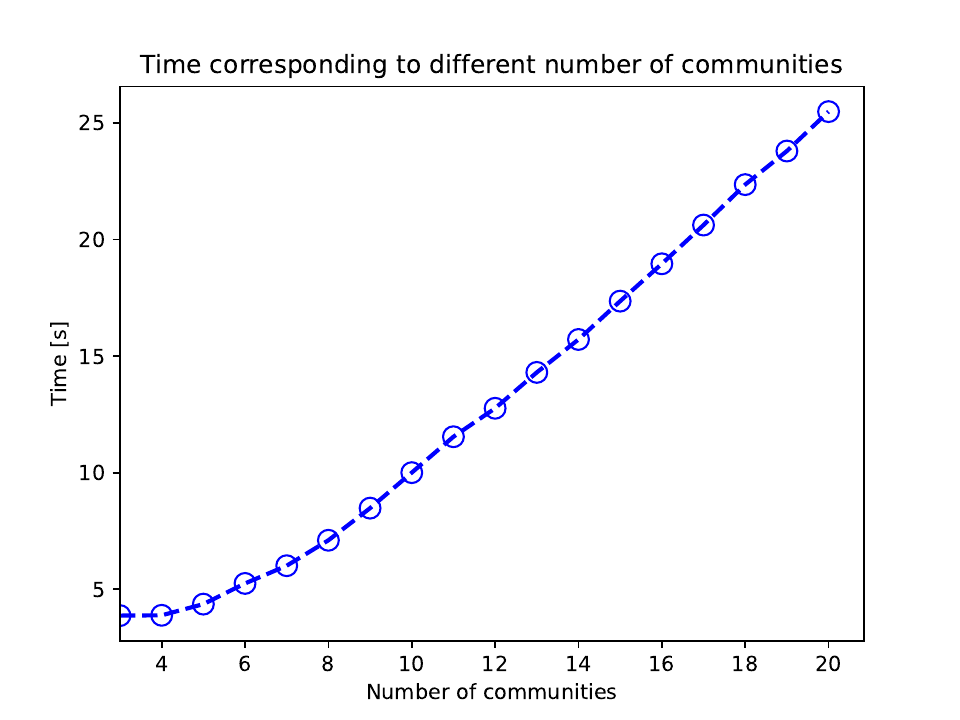}
     \caption{Time calculated by \DIFdelbeginFL \DIFdelFL{$\mathrm{APWP}$ }\DIFdelendFL \DIFaddbeginFL \DIFaddFL{$\mathrm{FPWP}$ }\DIFaddendFL with different numbers of communities. }
    \label{fig:N_cluster}
\end{figure}

In Figure~\ref{fig:N_cluster}, we plot the execution time of \DIFdelbegin \DIFdel{$\mathrm{APWP}$ }\DIFdelend \DIFaddbegin \DIFadd{$\mathrm{FPWP}$ }\DIFaddend depending on the number of communities. \rev{For this experiment we generated a random probabilistic network with 100 nodes and 500 edges. We vary the number of communities from 3 to 20, randomly assigning the nodes to them.}  From this figure, we observe that as the number of communities increases, the execution time of \DIFdelbegin \DIFdel{$\mathrm{APWP}$ }\DIFdelend \DIFaddbegin \DIFadd{$\mathrm{FPWP}$ }\DIFaddend also increases.

\begin{figure}[ht]
    \begin{center}
      \centerline{\includegraphics[width=.5\textwidth]{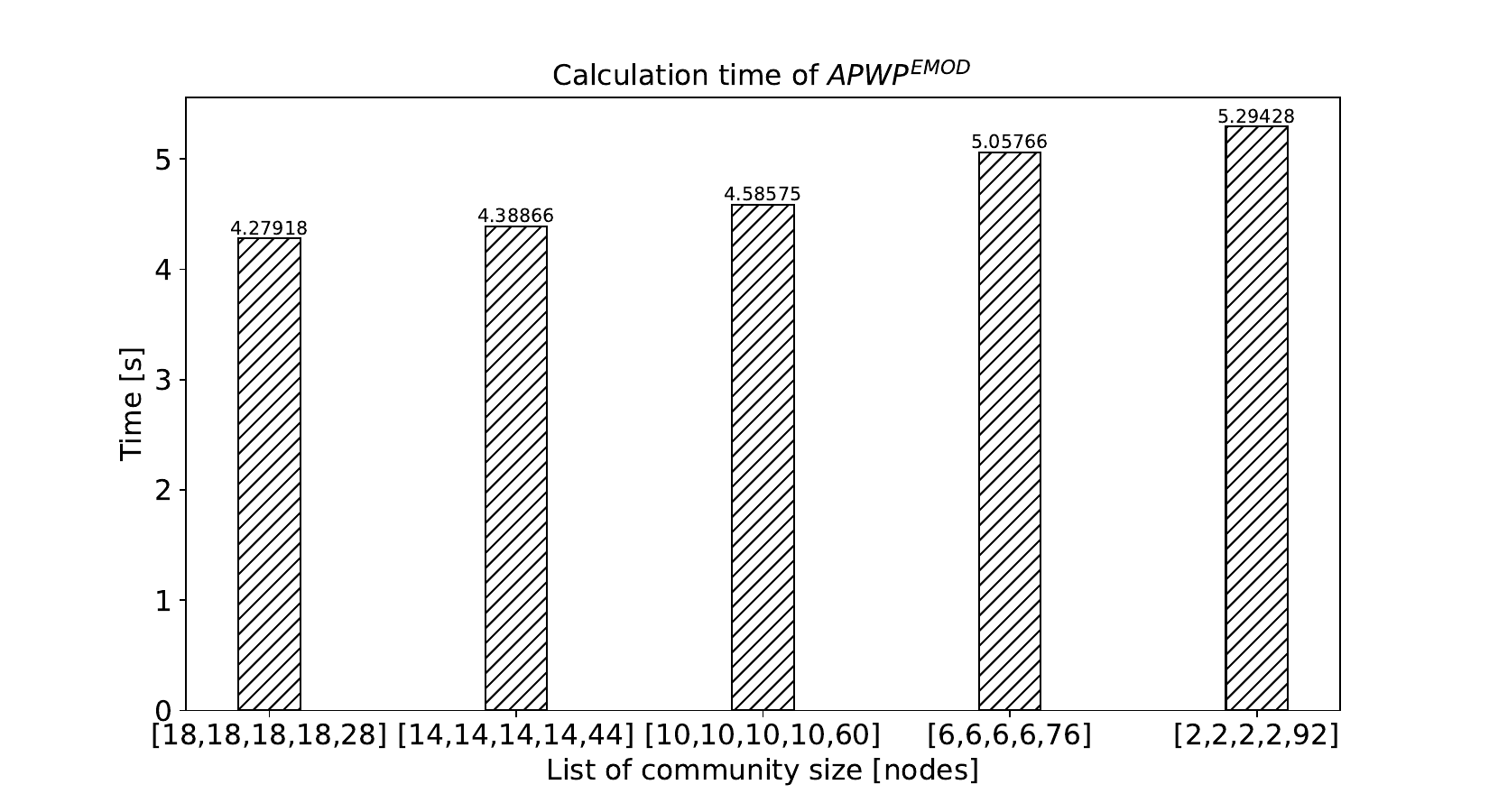}}
     \caption{Time calculated by \DIFdelbeginFL \DIFdelFL{$\mathrm{APWP}$ }\DIFdelendFL \DIFaddbeginFL \DIFaddFL{$\mathrm{FPWP}$ }\DIFaddendFL with different variances of communities size.}
    \label{fig:size_cluster}
    \end{center}
\end{figure}

In Figure~\ref{fig:size_cluster}, we show the performance of \DIFdelbegin \DIFdel{$\mathrm{APWP}$ }\DIFdelend \DIFaddbegin \DIFadd{$\mathrm{FPWP}$ }\DIFaddend with the same number of communities but different community size distributions. From this figure, we can find that the larger variance among communities, the longer the running time.

%\subsubsection*{Performance in different random networks }
In our final experiment, we  \rev{examine the execution time of} \DIFdelbegin \DIFdel{$\mathrm{APWP}$ }\DIFdelend \DIFaddbegin \DIFadd{$\mathrm{FPWP}$ }\DIFaddend on different random networks. In particular, we use 5 random network models: a Forest Fire Network (FFN) \cite{leskovec2005graphs}, Barabási-Abert (BA) \cite{barabasi1999emergence}, Small world (SW) \cite{watts1998collective}, Erdős-Rényi (ER) \cite{erdHos1960evolution}, and a clear community-structure network (\emph{CCS network}). We use each of these models to generate three random networks, all with around 200 nodes and 600 edges, but different numbers of communities. To provide a community assignment for the computation of modularity, we use a modularity optimization on the original deterministic networks. We also control the number of communities, where $|C|=4$, $|C|=5$ and $|C|=6$ respectively. 

\begin{figure}[ht]
    \centering
\begin{subfigure}[b]{0.24\textwidth}
         \centering
         \includegraphics[width=\textwidth]{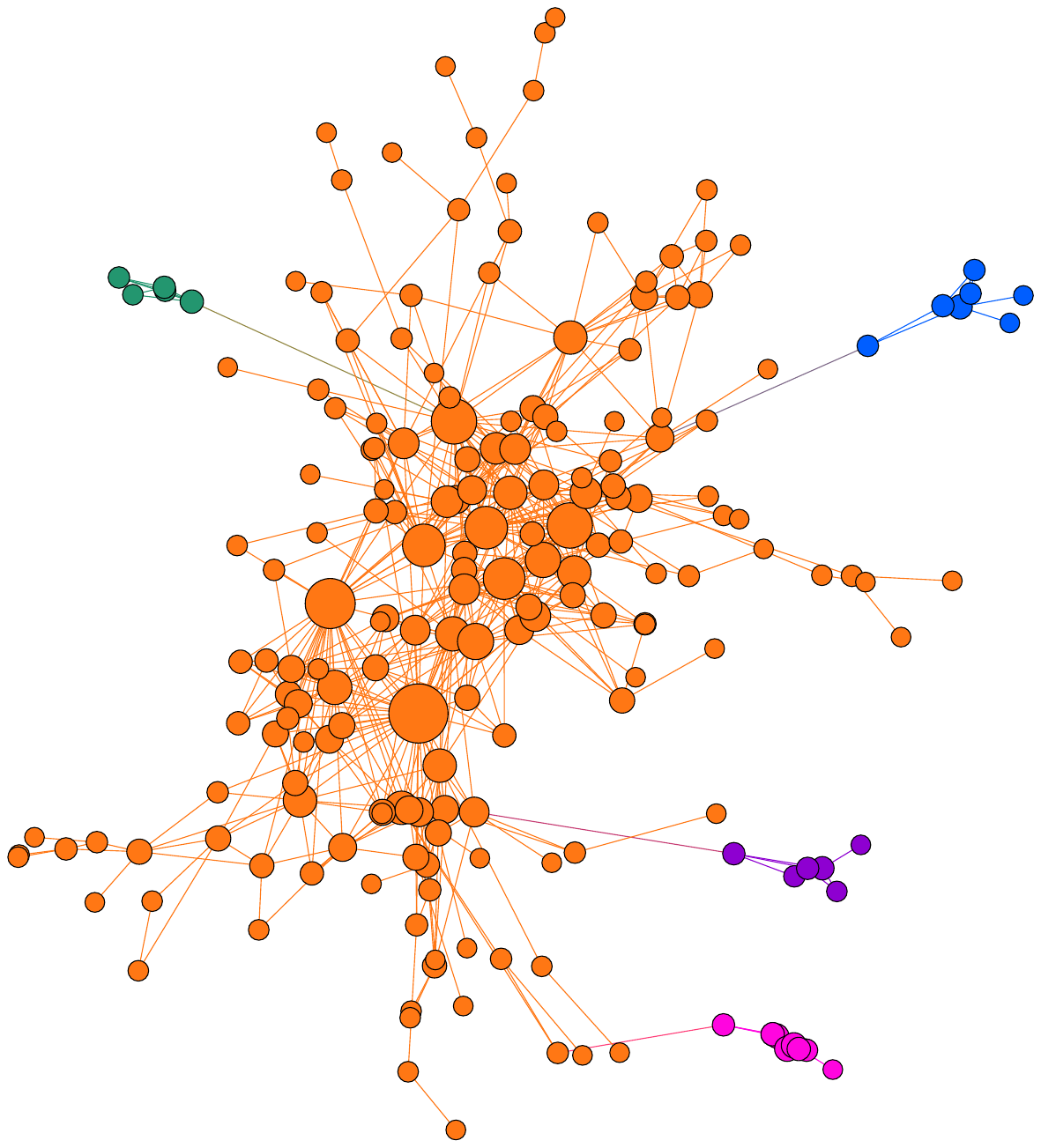}
         \caption{Forest Fire network}    
            \label{fig:ffn}
     \end{subfigure}
\begin{subfigure}[b]{0.24\textwidth}
         \centering
         \includegraphics[width=\textwidth]{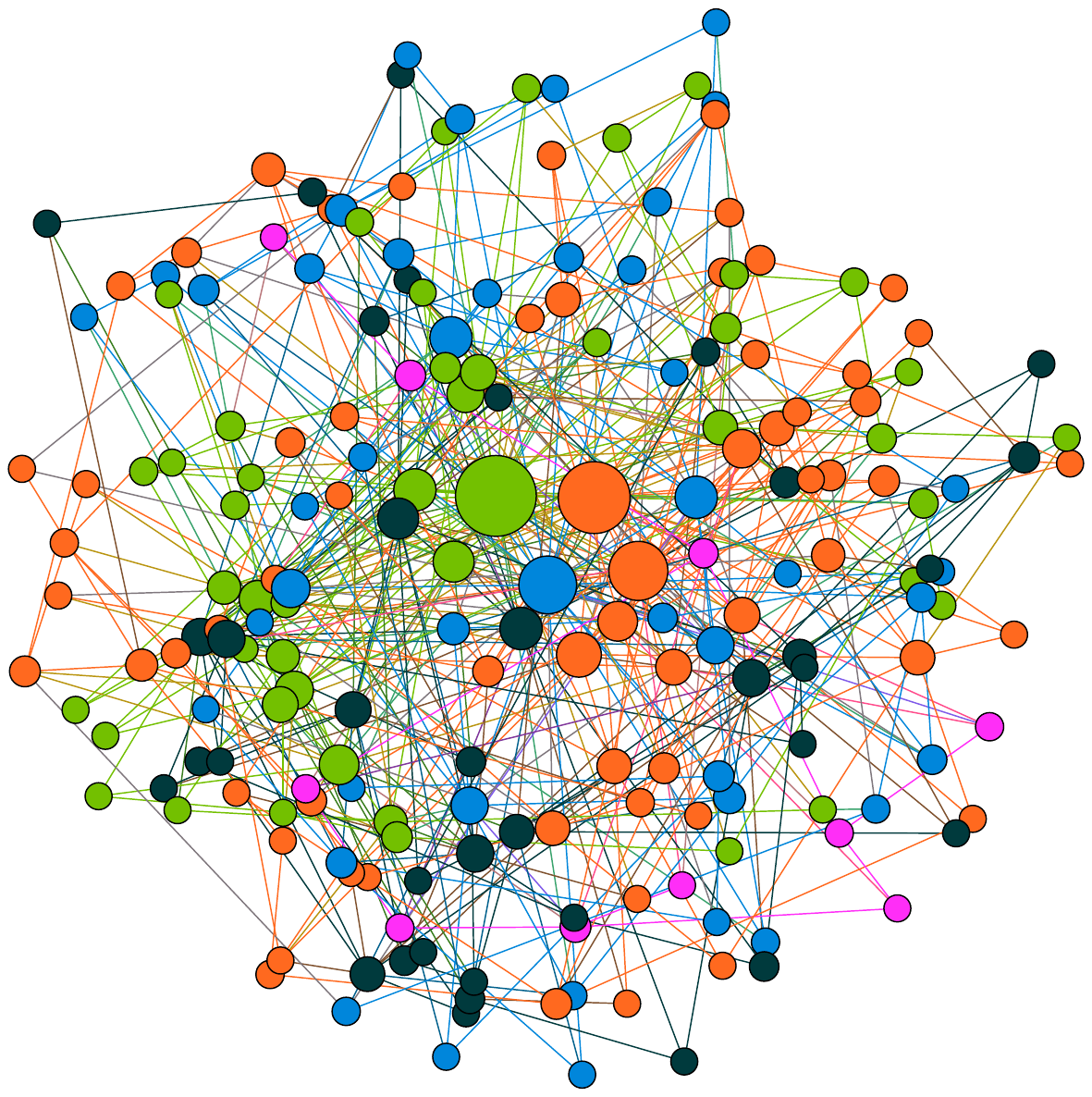}
         \caption{Barabási-Abert network}    
            \label{fig:ba}
     \end{subfigure}
\begin{subfigure}[b]{0.24\textwidth}
         \centering
         \includegraphics[width=\textwidth]{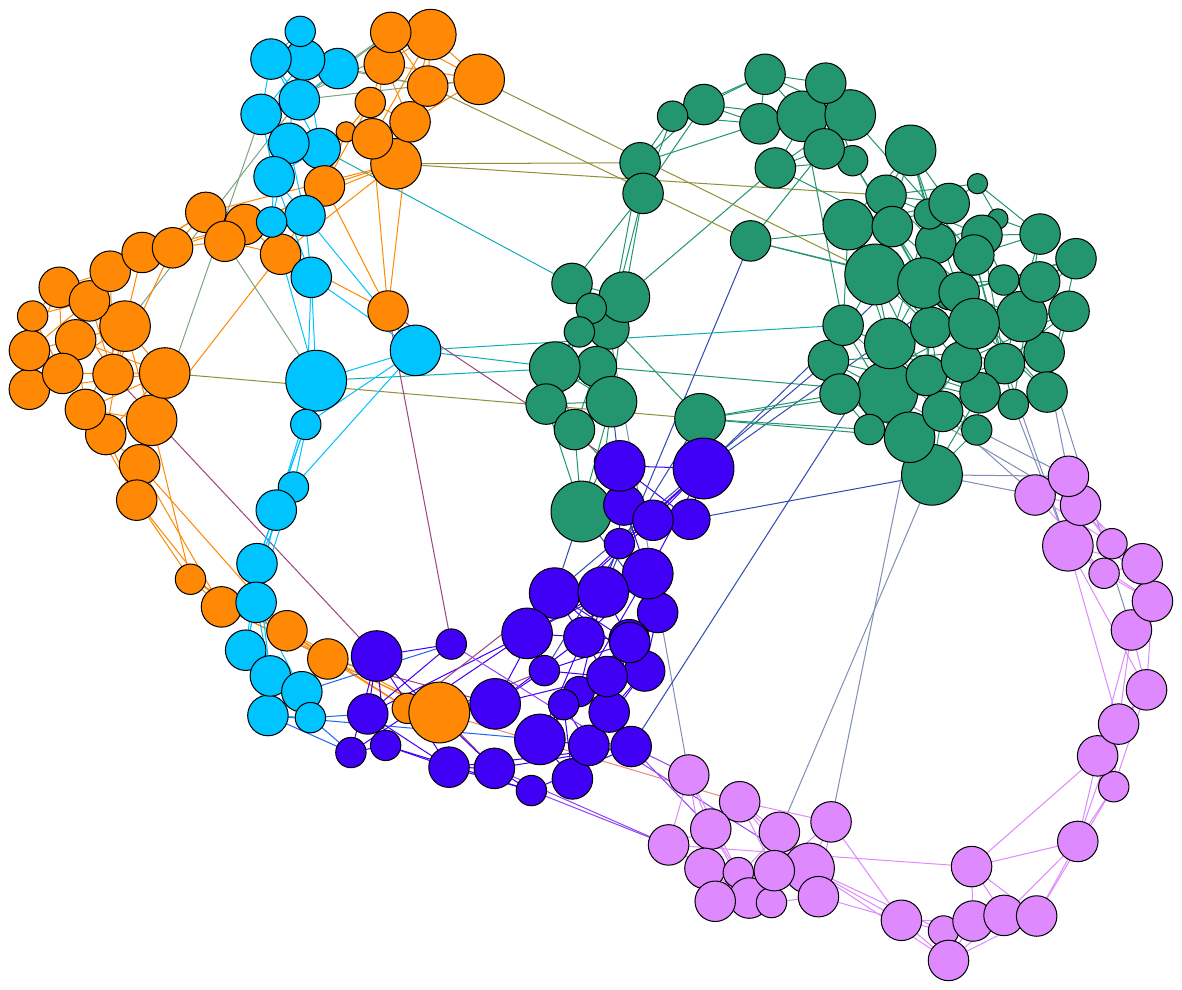}
         \caption{Small world network}    
            \label{fig:sw}
     \end{subfigure}
\begin{subfigure}[b]{0.24\textwidth}
         \centering
         \includegraphics[width=\textwidth]{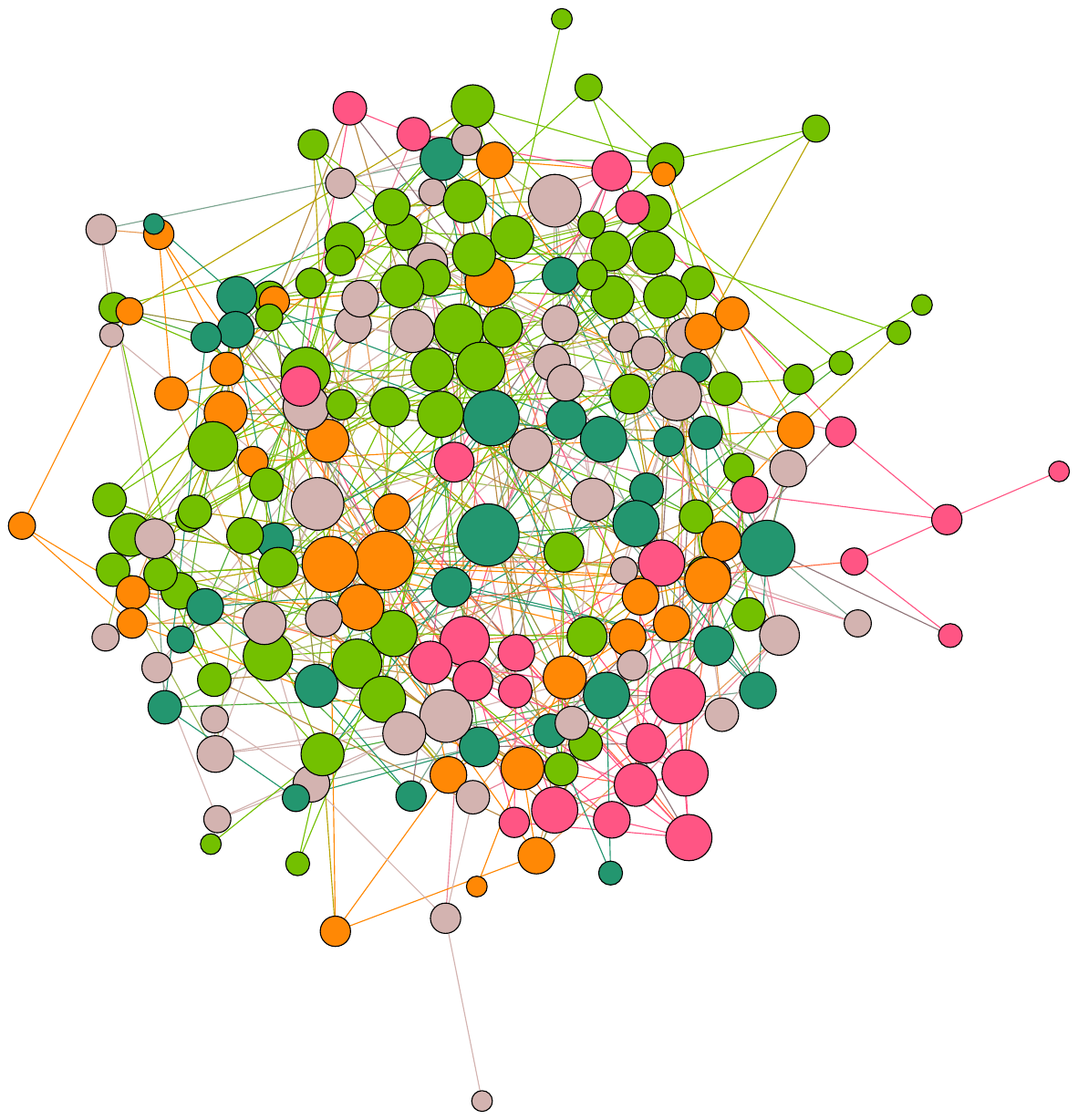}
         \caption{Erdős-Rényi network}    
            \label{fig:er}
     \end{subfigure}

\caption  {Networks generated by different models with 5 communities.} 
\label{fig:model_cluster}
\end{figure}

Figure~\ref{fig:model_cluster} shows different network structures containing 5 %\mm{Sometimes you write \emph{5}, sometimes \emph{five}. Be uniform (and check some writing guidelines to choose)} 
communities. Here nodes of the same color are in the same community computed using a modularity maximization method, and the number of colors represents the number of communities. 
%The size of nodes corresponds to their degree. 
Based on different network structures, we assign random edge probabilities so that entropy ratio is 0.4.

\begin{figure}[ht]
    \centering
    \includegraphics[width=.5\textwidth]{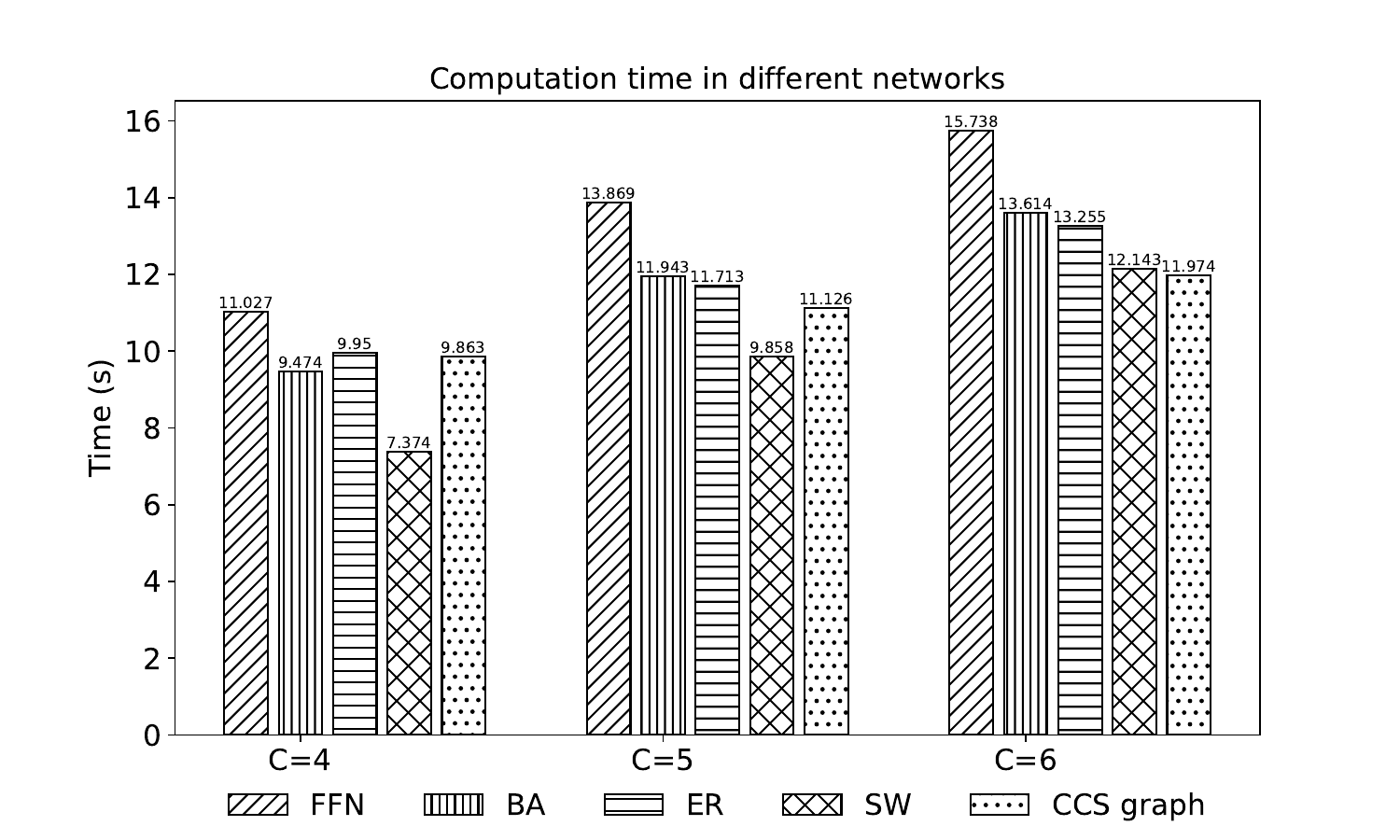}
     \caption{Running time of our method for different models.}
    \label{fig:com_time}
\end{figure}

Figure~\ref{fig:com_time} shows the running time of \DIFdelbegin \DIFdel{$\mathrm{APWP}$ }\DIFdelend \DIFaddbegin \DIFadd{$\mathrm{FPWP}$ }\DIFaddend on networks with different network structures and numbers of communities. Generally, with an increasing number of communities the running time also increases, which fits the results in Figure~\ref{fig:N_cluster}. %\mm{This is this section. Refer to the relevant Figure instead}. 
The visualization in Figure~\ref{fig:model_cluster} suggests that a modularity optimization method may find a larger central community in FFN, which  explains why the running time of FFN is always longer than for other networks. In fact, as shown in Figure~\ref{fig:size_cluster}, the larger the community size variance, the longer the running time.
%We can see that the running time of FFN is always longer than other networks. The visualization in \ref{fig:model_cluster} suggests that there exists one community that much larger than other 4 communities in FFN. And Figure~\ref{fig:size_cluster} explains that the larger size variance among communities, the longer time needed in calculation. 

\subsection*{Real data}
\begin{figure}[ht]
\centering
\begin{subfigure}{0.45\textwidth}
    \centering
    \includegraphics[width=\textwidth]{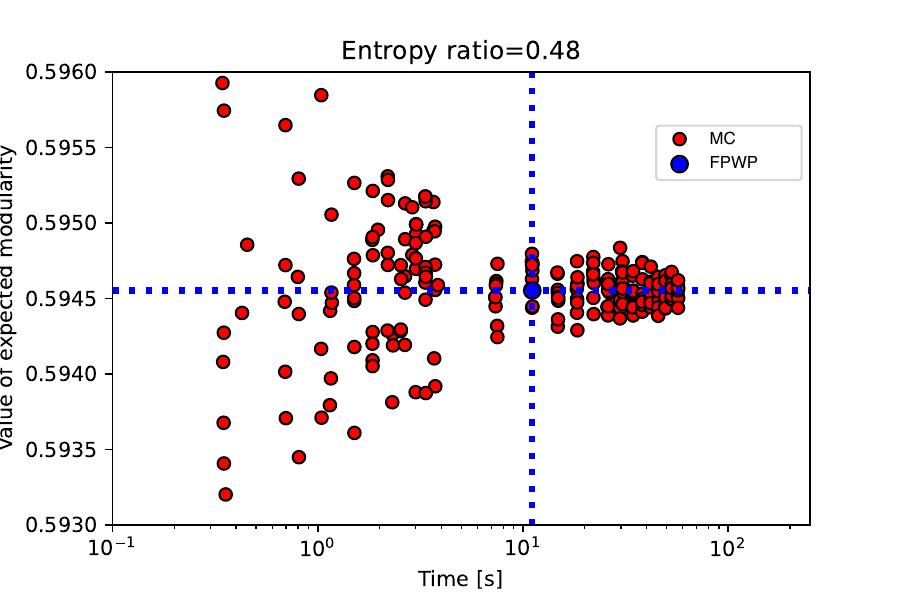}
    \caption{\rev{Enron network}}
    \label{fig:enron}
\end{subfigure}
\begin{subfigure}{0.45\textwidth}
    \centering
    \includegraphics[width=\textwidth]{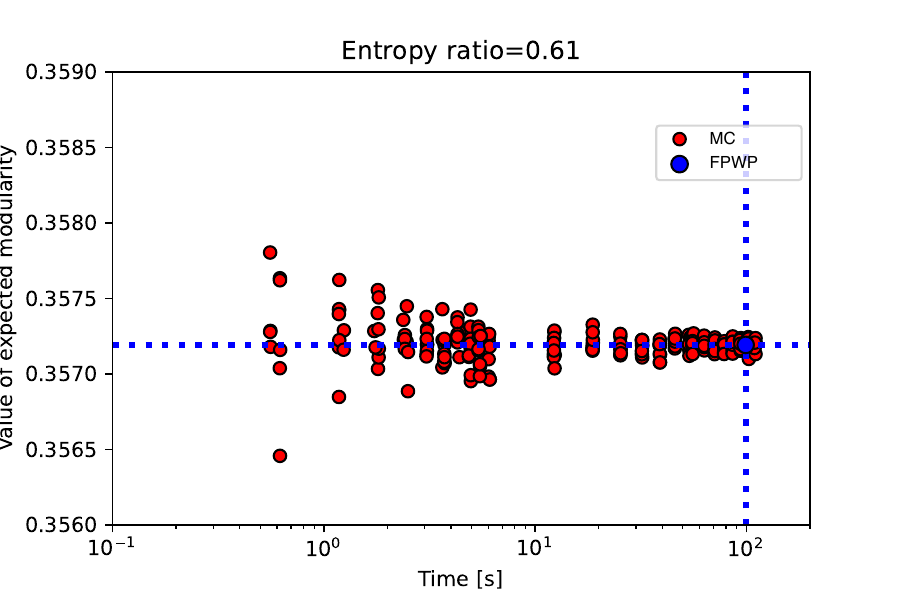}
    \caption{\DIFaddbegin \DIFadd{OSN}\DIFaddend}
    \label{fig:facebook}
\end{subfigure}\\
\begin{subfigure}{0.45\textwidth}
    \centering
    \includegraphics[width=\textwidth]{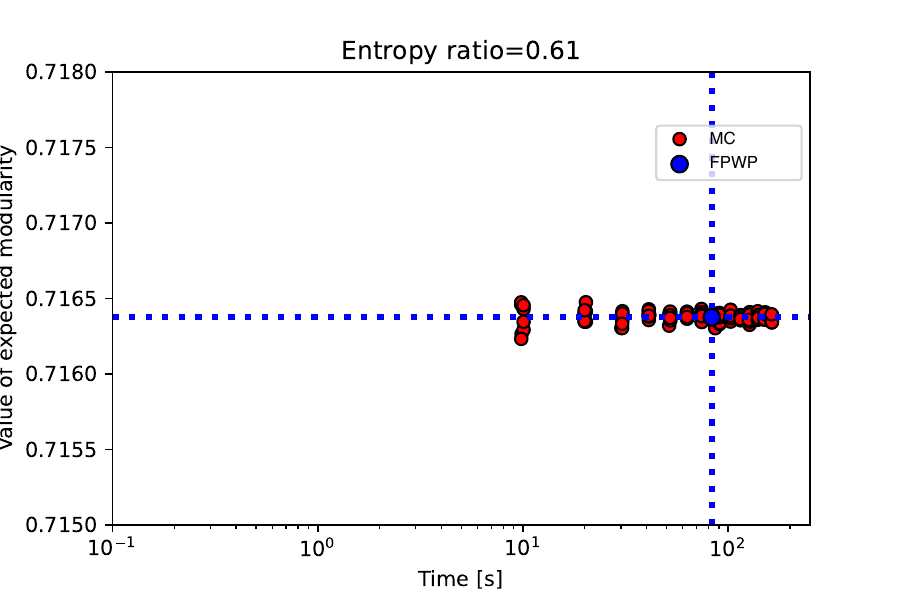}
      \caption{\rev{PPI network}}
    \label{fig:ppi}
\end{subfigure}
\begin{subfigure}{0.45\textwidth}
    \centering
    \includegraphics[width=\textwidth]{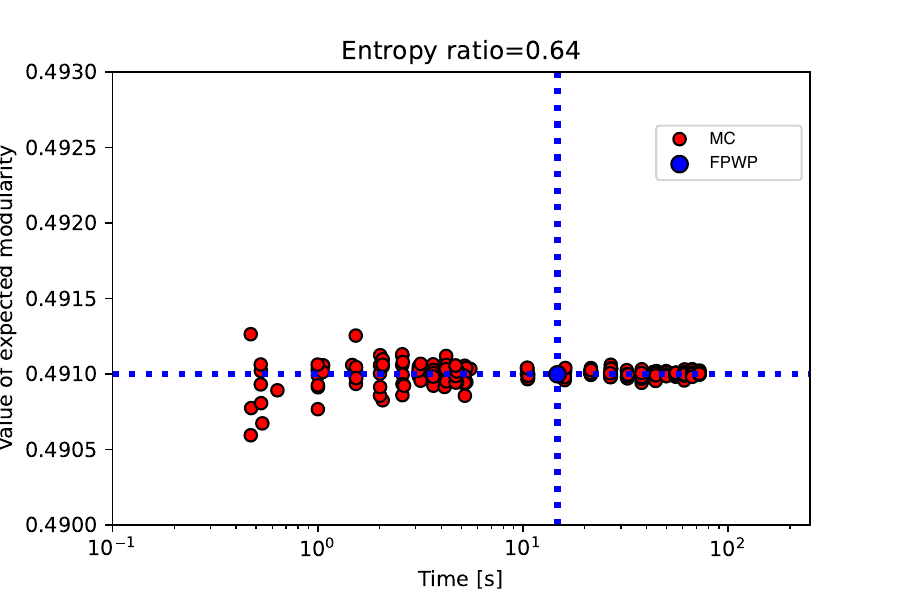}
     \caption{\rev{Collaboration network}}
    \label{fig:collaboration}
\end{subfigure}
\caption{Running time and expected modularity value calculated by different methods in real datasets.}
\label{fig:realdataset}
\end{figure}

\begin{figure}[ht]
\begin{subfigure}{0.24\textwidth}
    \centering
    \includegraphics[width=\textwidth]{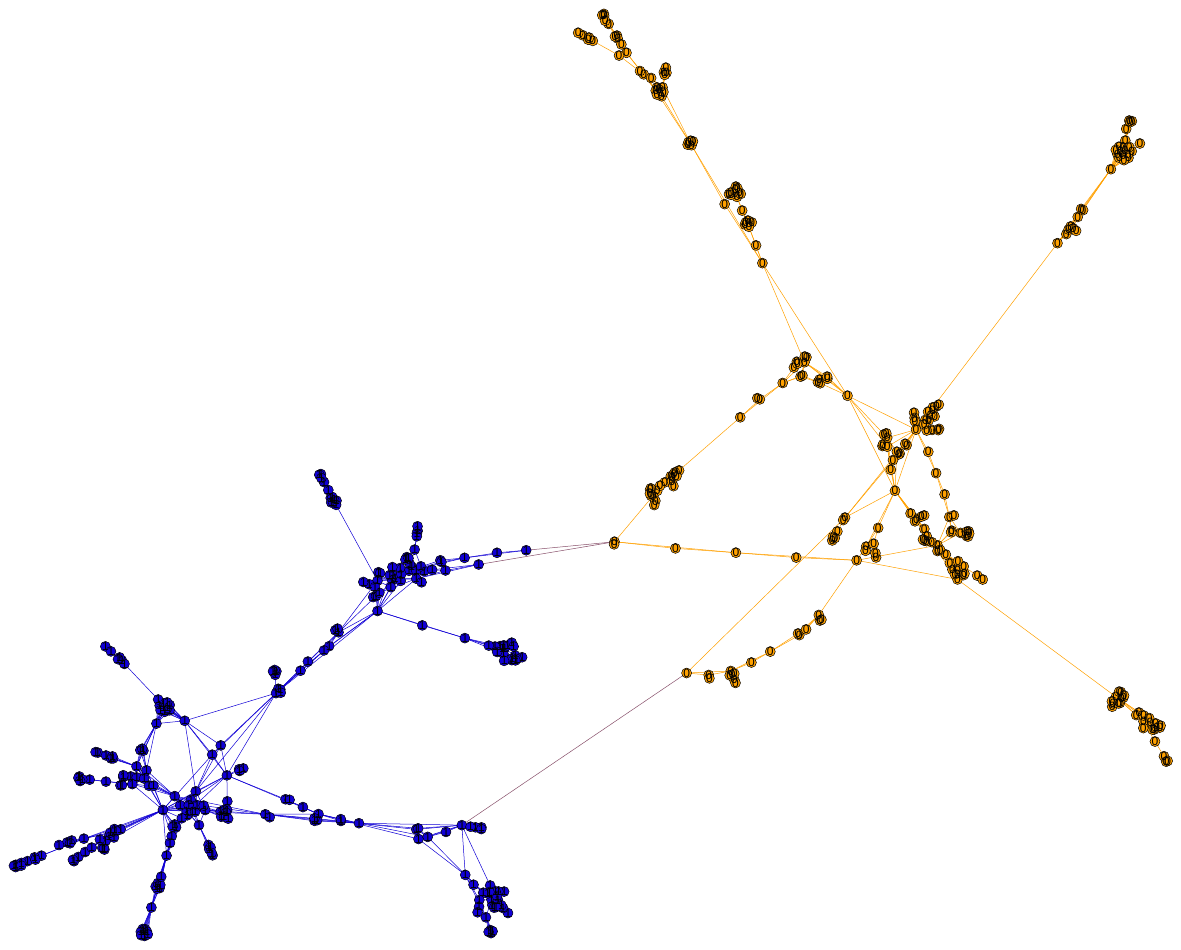}
     \caption{Collaboration network $|E|=1224, |C|=2$}
    \label{fig:collaborationcol}
\end{subfigure}    
\begin{subfigure}{0.24\textwidth}
    \centering
    \includegraphics[width=\textwidth]{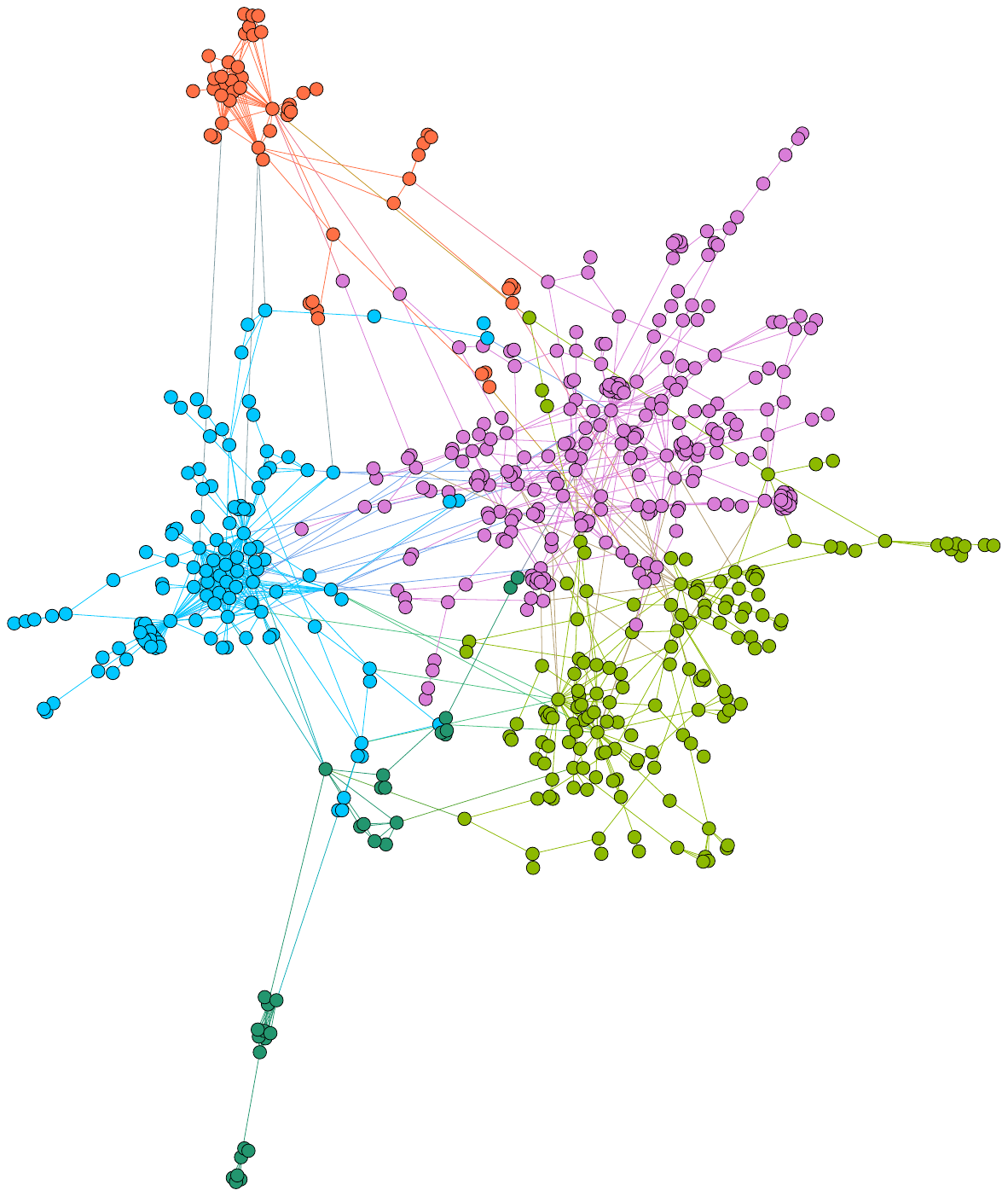}
     \caption{PPI network $|E|=1185, |C|=5$}
    \label{fig:collaborationppi}
\end{subfigure}
\begin{subfigure}{0.24\textwidth}
    \centering
    \includegraphics[width=\textwidth]{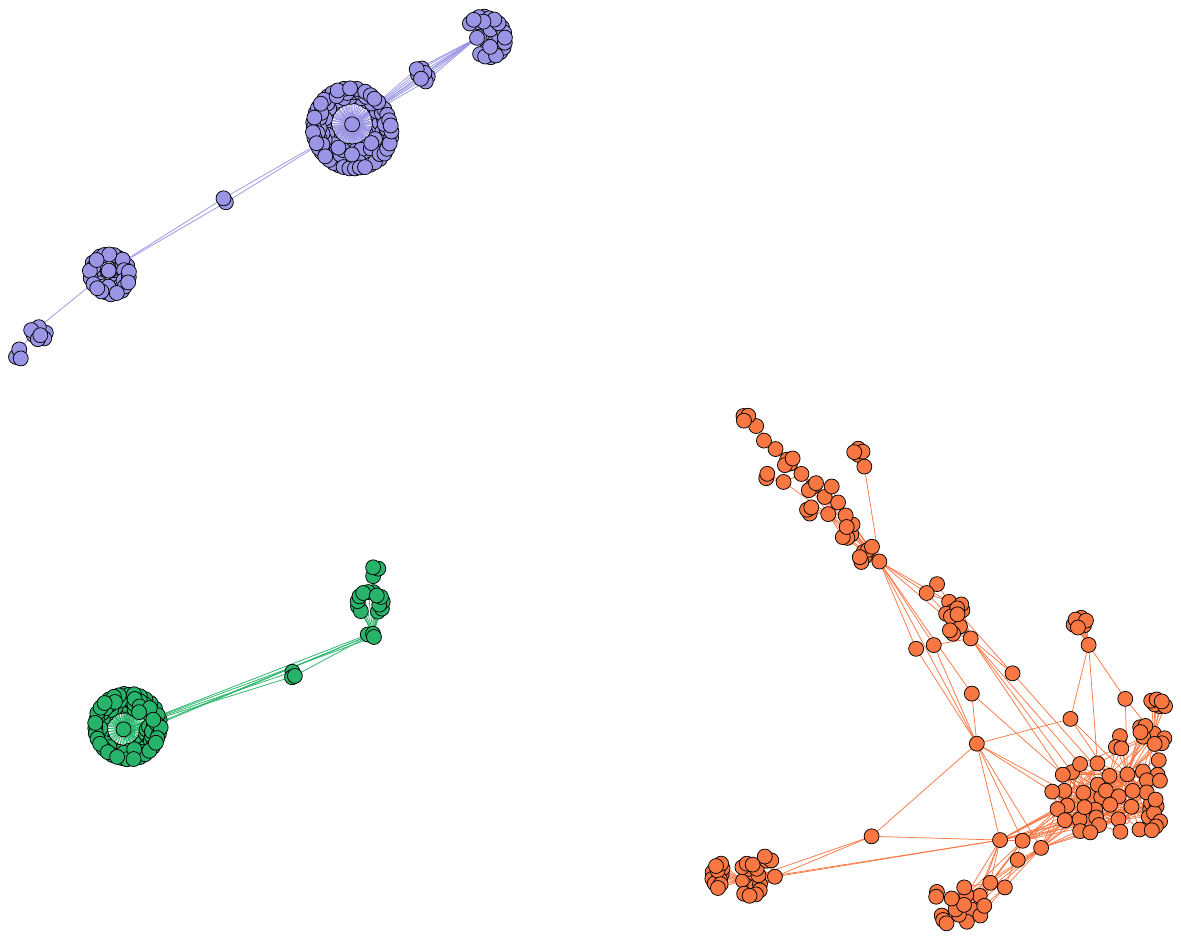}     
    \caption{Enron network $|E|=833, |C|=3$}
    \label{fig:collaborationenr}
\end{subfigure}  
\begin{subfigure}{0.24\textwidth}
    \centering
    \includegraphics[width=\textwidth]{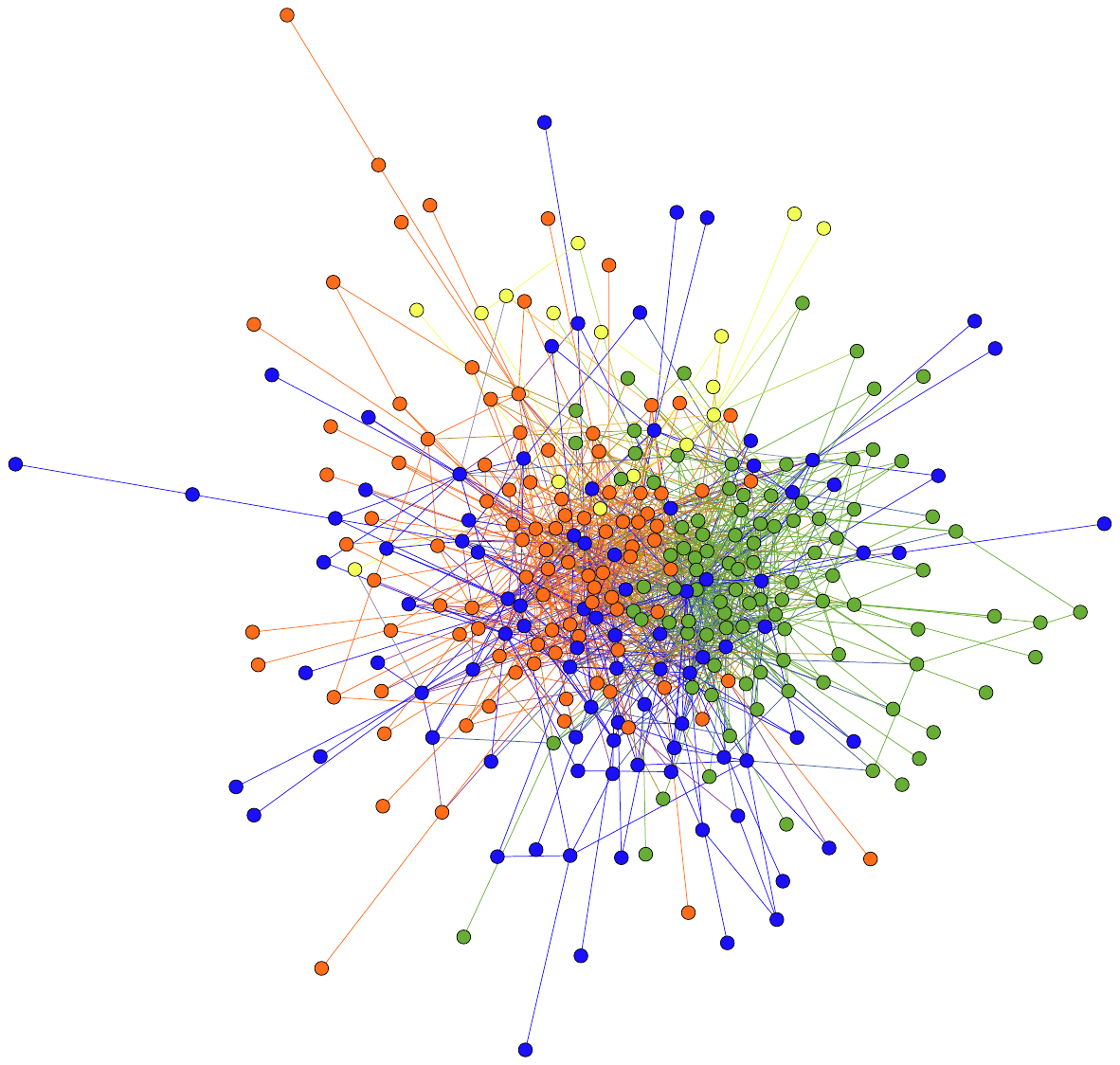}     
    \caption{\DIFaddbegin \DIFadd{OSN }\DIFaddend \\ $|E|=1217, |C|=4$}
    \label{fig:collaborationfac}
\end{subfigure}  

\caption{Real networks and the communities used in the experiments.}
\label{fig:real_visualization}
\end{figure}

\DIFdelbegin \DIFdel{\rev{In this section, we apply $\mathrm{APWP}$   and the sampling method on real networks. Notice that we do not use thresholding and weighting, because the experiments on synthetic data have clearly shown that these approaches should not be used to compute expected modularity. Figure~\ref{fig:real_visualization} shows different network structures with different numbers of clusters. Here nodes of the same color are in the same community computed using a modularity maximization method, and the number of colors represents the number of communities.  Figure~\ref{fig:realdataset} shows the execution time (vertical blue line) and the computed expected modularity (blue horizontal line) for our algorithm. Each red circle is the result of a different execution of the sampling method. For different executions, we have let the sampling method run for different amounts of time, so that we could inspect the balance between execution time and the likelihood of returning an accurate result for the sampling method. Figure~\ref{fig:realdataset} shows the same trends seen in the experiments with synthetic data. In these datasets, where the probability distributions are fixed and where clear community structures exist (we remind the reader that real data was obtained from high-modularity networks), the sampling method can produce  accurate results in a short time, in accordance with our results on synthetic data. %But if we don't have blue horizontal line, that is expected modularity value calculated by our method, it is hard to know how many samples we choose can get proper results when using sampling. This means that $\mathrm{APWP}$ can give the criteria that when sampling method can get the proper results, and when it could be stopped if people want to use sampling method.
} 
}\rev{In this section, we apply $\mathrm{FPWP}$   and the sampling method on real networks. Notice that we do not use thresholding and weighting, because the experiments on synthetic data have clearly shown that these approaches should not be used to compute expected modularity. Figure~\ref{fig:real_visualization} shows different network structures with different numbers of clusters. Here nodes of the same color are in the same community computed using a modularity maximization method, and the number of colors represents the number of communities.  Figure~\ref{fig:realdataset} shows the execution time (vertical blue line) and the computed expected modularity (blue horizontal line) for our algorithm. Each red circle is the result of a different execution of the sampling method. For different executions, we have let the sampling method run for different amounts of time, so that we could inspect the balance between execution time and the likelihood of returning an accurate result for the sampling method. Figure~\ref{fig:realdataset} shows the same trends seen in the experiments with synthetic data. In these datasets, where the probability distributions are fixed and where clear community structures exist (we remind the reader that real data was obtained from high-modularity networks), the sampling method can produce  accurate results in a short time, in accordance with our results on synthetic data. %But if we don't have blue horizontal line, that is expected modularity value calculated by our method, it is hard to know how many samples we choose can get proper results when using sampling. This means that $\mathrm{FPWP}$ can give the criteria that when sampling method can get the proper results, and when it could be stopped if people want to use sampling method.
}

 \section*{Discussion}

In this paper, we evaluate alternative approaches for the calculation of expected modularity in probabilistic networks, including an original approach compatible with possible worlds semantics. % but achieving polynomial complexity. %Our first method can be efficiently approximated without significantly affecting accuracy. %, by establishing a connection between calculation of partition's possibility and discrete Fourier transform calculation.

%We test our algorithm on different random networks against the main available alternative methods. Our experimental evaluation demonstrates that our algorithm not only produces accurate results, but also runs faster and is more robust than existing methods under different experimental conditions.
When compared with a brute-force approach, our new approach is several orders of magnitude faster, with \DIFdelbegin \DIFdel{$\mathrm{APWP}$ }\DIFdelend \DIFaddbegin \DIFadd{$\mathrm{FPWP}$ }\DIFaddend %\mm{in the paper are we saying \emph{approximate}, \emph{approximated}, or \emph{approximation} method. Make this uniform (and correct, in case one term is the correct one --- please check)} 
%method 
reducing time complexity from exponential to polynomial without 
%significantly 
affecting accuracy. \rev{For a network with 25 edges, which is around the limit of what we could handle using exact computation methods before this paper, the execution time %using our Python implementation 
goes down from more than 15 hours (brute-force) to less than 5 seconds ($\mathrm{PWP}$) to a few microseconds ($\mathrm{FPWP}$). }%While this enables the ca of expected modularity in other algorithms, e.g. to identify communities, t
The execution time is however still high for large networks. %, taking several minutes for networks with thousands of edges. %\mm{Is this correct? It's difficult to see the precise execution times from the log-log plot.}.
In cases when the computation has to be repeated multiple times sequentially, for example inside unoptimized expected modularity maximization algorithms, one may have to consider using faster but less accurate approaches.

%Our method is in general preferable to other common approaches used to handle probabilistic networks. 
We show that considering probabilities as weights, which can be a useful approach for other tasks, should not be used to compute expected modularity because it returns wrong results in general. The main reason is that node strengths (which equal node expected degrees in probabilistic networks) are multiplied in weighted modularity computation, while the expected product of degrees is not the same as the product of expected degrees in case of dependencies. Unfortunately, degrees in probabilistic networks are not independent, even when edge probabilities are, because edges are incident to two nodes and thus they either contribute to the degree of both or of none of them. A second reason is that weights are interpreted relative to each other in weighted modularity. So, for example, having all weights set to 0.01 or having them set to 1 makes no difference, while interpreting these numbers as probabilities means that in one case the network is probably empty, while in the second case all edges are almost certainly there. Identifying cases where using weighted modularity instead of expected modularity can still be appropriate is an interesting research question raised by this work.

Thresholding is a common and very fast approach to handle probabilistic networks. In fact, we argue that even if the large majority of network studies currently use deterministic networks, these are in fact often uncertain networks where some more or less explicit decisions have been made about including or not edges based on the available information about the modeled system --- that is, some thresholding has been performed. In this paper we show that thresholding is not a good approach for the task of computing expected modularity in general. However this is the case only when the network is almost deterministic. If we assume that all probabilities are at distance $\epsilon$ from 0 or 1, using thresholding we change edge probabilities from $1 - \epsilon$ to 1 and from $\epsilon$ to 0, for all thresholds between $\epsilon$ and $1 - \epsilon$. With small values of $\epsilon$, the change is small and the range of good thresholds is wide, making the method fast, accurate, and robust. However, as we are interested in handling probabilistic networks, it is not very useful to have a method that only works when the network is practically not probabilistic.

%A similar consideration holds for sampling: s
While the amount of uncertainty has practically no effect on the accuracy and execution time of the new method we introduce in this paper, this is an important factor when sampling is used. This feature of sampling is already well acknowledged in the literature, where some methods have been proposed to tranform the original network into another probabilistic network of lower entropy \cite{parchas2018uncertain,kaveh_probabilistic_2021}. Sampling from a probabilistic network with probabilities very close to 0 or 1 immediately gives an accurate result, the same computed on the thresholded deterministic network. We have also noticed how the likelihood of producing an accurate value of expected modularity using sampling increases when the community structure is very clear, close to a set of cliques. This is due to the fact that modularity depends on the number of edges inside communities, and not on the position of these edges. As a result, when many (and most) edges are inside communities, many sampled networks will have a similar number of edges inside those communities, and thus similar values of modularity. Sampling from a narrow distribution has then a faster convergence. 
However, %in most interesting cases our algorithm is preferable to sampling: it returns an accurate value whose accuracy does not depend on the execution time, while 
sampling can produce an inaccurate result, in addition to not giving any indication of whether the result is accurate or not. The more the network is uncertain, the less well sampling works.
%, while our algorithms are not affected by the probability distribution (neither with respect to accuracy nor execution time).

While the probability distribution does not impact the execution time of our method, the way in which nodes are assigned to communities does, both with respect to the number of communities and to the distribution of community sizes. The experiments on different types of networks (with or without a long-tail degree distribution, with or without a high clustering coefficient) also show that our algorithms are not significantly and directly affected by the network type, although depending on the network type it can be more or less likely to have to execute our algorithm with large communities, which would then indirectly affect execution time.

This paper opens the problem of how to use 
%our algorithm 
methods for the computation of expected modularity
as a sub-procedure of a community detection method. Such an algorithm has not been proposed yet, despite the popularity of its deterministic counterpart. So far, clustering algorithms for probabilistic networks have been mostly developed outside of the network science research community, overlooking the popularity of the modularity objective function\cite{kollios2011clustering, qiu_efficient_2019, han_efficient_2019, liang_efficient_2020}.
This problem adds two interesting aspects to our work: first, when used as a heuristic inside an optimization algorithm, a less accurate calculation of modularity may still lead to the same clustering, so there is a question about when it is needed to be accurate and when a faster approximation is sufficient. That is, both sampling and considering probabilities as weights could be usable inside a community detection algorithm, even if their results may be inaccurate. Second, we may have to execute the modularity calculation many times, which poses an additional computational challenge in particular for large networks. These two aspects may allow the usage of different computational approaches\cite{qiao2018fast,tian2014learning}.

%DIF < \rev{Our work could be regarded as the basis of community detection based on modularity. It involves running our algorithm $APWP^{EMOD}$ many times, leading to the challenge of optimizing a sequence of execution, which we suggest for future work.}
%DIF > \rev{Our work could be regarded as the basis of community detection based on modularity. It involves running our algorithm $FPWP^{EMOD}$ many times, leading to the challenge of optimizing a sequence of execution, which we suggest for future work.}

%\bibliographystyle{naturemag-doi}
\bibliography{references}

\section*{Acknowledgments}

This work has been partly funded by eSSENCE, an e-Science collaboration funded as a strategic research area of Sweden, by the Centre of Natural Hazards and Disaster Science (CNDS), by the Wallenberg AI, Autonomous Systems and Software Program WASP, and by the AI4Research initiative at Uppsala University.

\section*{Author contributions statement}
% Categories from: https://www.elsevier.com/researcher/author/policies-and-guidelines/credit-author-statement
Data Curation, Software, Investigation, Writing -- Original Draft: X.S.; 
Methodology: X.S., M.M., C.R.
Conceptualization, Writing -- Review \& Editing: X.S., M.M., C.R., F.S.; Supervision, Funding acquisition: M.M., C.R.

%X.S. developed the initial idea, conducted the majority of the theoretical analysis and experiments, and drafted the original manuscript. M.M. contributed to refining the theoretical analysis, assisted with experimental design, and helped revise the manuscript. C.R. provided useful feedback on the analysis and experiments and assisted with revisions. F.S. gave useful input on the theoretical analysis and helped write the final manuscript
%X.S. led all aspects of the work, X.S., M.M., C.R., F.S. designed the methods, X.S., M.M., C.R. conceived the experiments, X.S. conducted the experiments, X.S., M.M., C.R., F.S. analysed the results.  All authors reviewed the manuscript. 
\section*{Data availability statement}
The datasets generated and analysed during the current study are available in the repository: \href{https://github.com/XINS3/Expected-modularity-calculation-over-uncertain-graph}{https://github.com/XINS3/Expected-modularity-calculation-over-uncertain-graph}. 
\section*{Additional information}

The code to replicate the experiments is available here: \href{https://github.com/XINS3/Expected-modularity-calculation-over-uncertain-graph}{https://github.com/XINS3/Expected-modularity-calculation-over-uncertain-graph}; the authors declare no competing interests. 
% https://github.com/uuinfolab/Expected\_Modularity\_Calculation\_in\_ProbabilisticGraph

\end{document}